\documentclass[prd,nofootinbib,preprint,superscriptaddress]{revtex4}
\pdfoutput=1

\usepackage{amsmath,amssymb}
\usepackage{epsfig}
\usepackage{graphicx}
\usepackage[usenames,dvipsnames]{color}
\usepackage{subfigure}
\usepackage{slashed}
\usepackage[colorlinks,citecolor=blue]{hyperref}
\usepackage{color}
{
	{
		
\begin{document}
\title{Self-interacting Inelastic Dark Matter in the Light of XENON1T Excess}
			
\author{Manoranjan Dutta}
\email{ph18resch11007@iith.ac.in}
\affiliation{Department of Physics, Indian Institute of Technology Hyderabad, Kandi, Sangareddy 502285, Telangana, India}
			
\author{Satyabrata Mahapatra}
\email{ph18resch11001@iith.ac.in}
\affiliation{Department of Physics, Indian Institute of Technology Hyderabad, Kandi, Sangareddy 502285, Telangana, India}
			
\author{Debasish Borah}
\email{dborah@iitg.ac.in}
\affiliation{Department of Physics, Indian Institute of Technology Guwahati, Assam 781039, India}
			
\author{Narendra Sahu}
\email{nsahu@phy.iith.ac.in}
\affiliation{Department of Physics, Indian Institute of Technology Hyderabad, Kandi, Sangareddy 502285, Telangana, India}
			
\begin{abstract}
We propose a self-interacting inelastic dark matter (DM) scenario as a possible origin of the recently reported excess of electron recoil events by the XENON1T experiment. Two quasi-degenerate Majorana fermion DM interact within themselves via a light hidden sector massive gauge boson and with the standard model particles via gauge kinetic mixing. We also consider an additional long-lived singlet scalar which helps in realising correct dark matter relic abundance via a hybrid setup  comprising of both freeze-in and freeze-out mechanisms. While being consistent with the required DM phenomenology along with sufficient self-interactions to address the small scale issues of cold dark matter, the model with GeV scale DM can explain the XENON1T excess via inelastic down scattering of heavier DM component into the lighter one. All these requirements leave a very tiny parameter space keeping the model very predictive for near future experiments.
\end{abstract}	
\maketitle
			
\section{Introduction}
\label{intro}
There exist convincing number of evidences suggesting the presence of a non-luminous, non-baryonic form of matter in the present universe, 
popularly known as dark matter (DM). This form of matter constitute a significant portion of galaxies, clusters as well as the entire universe. 
Data from cosmology experiments like Planck which measures the cosmic microwave background (CMB) anisotropies very precisely, predict the amount 
of DM in the present universe to be around $26.8\%$ of the present universe's energy density. In terms of density parameter $\Omega_{\rm DM}$ and 
$h = \text{Hubble Parameter}/(100 \;\text{km} ~\text{s}^{-1} 
\text{Mpc}^{-1})$, the present DM abundance is conventionally reported as \cite{Aghanim:2018eyx}:
$\Omega_{\text{DM}} h^2 = 0.120\pm 0.001$
at 68\% CL. Similar evidences exist in galactic and cluster scales as well, collected over a long period of time since 1930s \cite{Zwicky:1933gu, Rubin:1970zza, Clowe:2006eq}. It should be noted that the Planck estimate of present DM abundance relies upon the standard model of cosmology or ${\rm \Lambda CDM}$ cosmology which has been very successful in overall description of our universe at large scale $(\geq \mathcal{O}(\rm Mpc))$. Here CDM refers to cold dark matter while $\Lambda$ denotes the cosmological constant or dark energy. CDM, a pressure-less or collision-less fluid acts like a seed for structure formation providing the required gravitational potential well for ordinary matter to collapse and form structures. Since none of the standard model (SM) particles can be a viable CDM candidate, 
several beyond standard model (BSM) proposals have been put forwarded out of which the weakly interacting massive particle (WIMP) paradigm is the most widely studied one. In this framework, a WIMP candidate typically having interactions and mass in the electroweak regime, naturally satisfies the correct DM relic abundance, a remarkable coincidence often referred to as the \textit{WIMP Miracle}~\cite{Kolb:1990vq}.

While ${\rm \Lambda CDM}$ is in excellent agreement with large scale structure of the universe, yet there exist some discrepancies between its prediction and observations, particularly at small scales. In particular, too-big-to-fail, missing satellite and core-cusp problem are three such well known cases where 
${\rm \Lambda CDM}$ appears to be in conflict with observations. For recent reviews of these issues and possible solutions, please see \cite{Tulin:2017ara, Bullock:2017xww}. One interesting solution to this puzzle was proposed by Spergel and Steinhardt \cite{Spergel:1999mh} where they considered an alternative to collision-less CDM in terms of self-interacting dark matter (SIDM)\footnote{See \cite{deLaix:1995vi} for earlier studies.}. While SIDM solves the problems at small scales, it reproduces the CDM halos at large radii, thus consistent with observations. This is simply due to the fact that self-interacting scattering rate is proportional to DM density. The required self-interaction rate is often quantified as a ratio of cross section to DM mass as $\sigma/m \sim 1 \; {\rm cm}^2/{\rm g} \approx 2 \times 10^{-24} \; {\rm cm}^2/{\rm GeV}$ \cite{Buckley:2009in, Feng:2009hw, Feng:2009mn, Loeb:2010gj, Zavala:2012us, Vogelsberger:2012ku}. Such self-interacting cross sections can be naturally realised in models with very light mediator. For such a scenario, self-interactions can be shown to be stronger for smaller DM velocities such that it can have large impact on small scale structures while being consistent with usual CDM predictions at larger scales \cite{Buckley:2009in, Feng:2009hw, Feng:2009mn, Loeb:2010gj, Bringmann:2016din, Kaplinghat:2015aga, Aarssen:2012fx, Tulin:2013teo}. From particle physics point of view, such self-interactions can be naturally realised in Abelian gauge extensions of the SM. While DM sector can not be completely hidden and there should be some coupling of the mediator with SM particles as well, which can ensure that DM and SM sectors were in thermal equilibrium in the early universe. The same coupling can also be probed at DM direct detection experiments as well \cite{Kaplinghat:2013yxa, DelNobile:2015uua}. Several model building efforts have been made to realise such scenarios. For example, see \cite{Kouvaris:2014uoa, Bernal:2015ova, Kainulainen:2015sva, Hambye:2019tjt, Cirelli:2016rnw, Kahlhoefer:2017umn} and references therein.

DM with light mediators have also received attention very recently after XENON1T collaboration published their latest results in June 2020 where they 
have reported the observation of an excess of electron recoil events over the background in the recoil energy $E_r$ in a range 1-7 keV, 
peaked around 2.4 keV\cite{Aprile:2020tmw}. While the excess can be explained by solar axions at $3.5\sigma$ significance or neutrinos with magnetic moment at $3.2\sigma$ significance both these interpretations face stringent stellar cooling bounds. While there is also room for possible tritium backgrounds in the detector, which XENON1T collaboration neither confirm or rule out at this stage, there have been several interesting new physics proposals in the literature. For example, see~\cite{Takahashi:2020bpq,Alonso-Alvarez:2020cdv,Kannike:2020agf,Fornal:2020npv,Du:2020ybt,Su:2020zny,Harigaya:2020ckz, Borah:2020jzi,Choudhury:2020xui, Bramante:2020zos, Bell:2020bes,  Borah:2020smw, Aboubrahim:2020iwb, Lee:2020wmh, Baek:2020owl, Shakeri_2020, Bally:2020yid, DelleRose:2020pbh, Ema:2020fit} and references therein. The DM interpretations out of these examples, typically have a light mediator via which DM interacts with electrons. The recoil can occur either due to light boosted DM or inelastic up or down-scattering~\cite{   Bell:2020bes, Lee:2020wmh, Baek:2020owl, Harigaya:2020ckz, Bramante:2020zos, Baryakhtar:2020rwy, Chao:2020yro, An:2020tcg, He:2020wjs, Choudhury:2020xui, Borah:2020jzi, Shakeri_2020, Borah:2020smw, Keung:2020uew, Aboubrahim:2020iwb, He:2020sat, Choi:2020ysq}.

Thus we noticed that in a class of models, the DM interpretation of XENON1T excess as well as SIDM phenomenology rely 
on light mediators. This motivates us to propose a common platform to show that the self interaction of DM arising via light mediators in 
such models can also give rise the observed XENON1T excess. In other words, the proposed scenario provides a unique way of probing the parameter 
space of SIDM at direct DM search experiments like XENON1T. To be more specific, we consider a dark sector consisting of sub-GeV inelastic DM 
with keV scale mass splitting and a corresponding massive vector boson $Z'$ \cite{Harigaya:2020ckz, Borah:2020smw}. Unlike earlier works where 
DM and $Z'$ masses are in the same regime so that DM relic is governed by resonant $2 \rightarrow 2$ annihilations, here we consider light 
mediators (order of magnitude lighter than DM mass) motivated from SIDM point of view. While the self interaction of DM is realised via $Z'$-exchange, 
the latter can mix with $U(1)_Y$ gauge boson to provide a unique portal for detecting the DM at direct search experiments. The scalar field which 
leads to spontaneous breaking of dark sector gauge symmetry also induces a tiny Majorana mass to a singlet Dirac fermion field leading to an 
inelastic DM scenario \cite{TuckerSmith:2001hy, Cui:2009xq}. In this setup we first find the DM parameter space consistent with velocity dependent 
self-interaction rates explaining the data at the scale of clusters, galaxies and dwarf galaxies. We then confront the SIDM parameter space with 
the observed XENON1T electron excess while being consistent with other experimental bounds. We show that these two requirements make pure thermal 
relic DM insufficient to produce the observed relic and therefore we consider a hybrid setup where both freeze-out and freeze-in mechanisms can 
play non-trivial roles in generating DM relic. As we discuss in the upcoming sections, a long lived scalar singlet has to be invoked whose late 
decay into DM helps in generating correct DM relic in such a hybrid setup.

This paper is organised as follows. In section \ref{sec2} we briefly discuss our model followed by the analysis for dark matter self-interaction in section \ref{sec3}. In section \ref{dmrelic}, we discuss production of self-interacting DM from a hybrid of freeze-in and freeze-out formalism. In section \ref{xenon1t} we discuss the possible origin of XENON1T excess in our model. We finally summarise our results and conclude in section \ref{conclude}.

\section{The Model}
\label{sec2}

We consider a simple Abelian extension of the SM. Under this $U(1)_X$ gauge symmetry, the SM fields do not have any charge while there exists 
a SM singlet Dirac fermion $\Psi$ with $U(1)_X$ charge $1$. A SM singlet scalar $\Phi$ having $U(1)_X$ charge $-2$ introduced which not only 
break the new gauge symmetry spontaneously but also splits the Dirac fermion into two pseudo-Dirac components as we discuss below. The Dirac fermion 
$\Psi$ is identified as the DM field. The relevant part of the DM Lagrangian is 
\begin{equation}
	\label{Lagrangian}
	\mathcal{L}_{DM} = i \overline{\Psi} \gamma^\mu D_\mu \Psi - M (\overline{\Psi}_L \Psi_R + \overline{\Psi}_R \Psi_L) -(y_L \Phi \overline{(\Psi_L)^c}\Psi_L) + (y_R \Phi \overline{(\Psi_R)^c}\Psi_R + h.c.)   + \frac{\epsilon}{2}B^{\alpha \beta}Y_{\alpha\beta}
\end{equation}
where $D_\mu = \partial_\mu + i g' Z'_\mu$ and $B^{\alpha\beta}, Y_{\alpha \beta}$ are the field strength tensors of $U(1)_X, U(1)_Y$ respectively and $\epsilon$ is the kinetic mixing between them. The Lagrangian involving singlet scalar can be written as
\begin{equation}
	\mathcal{L}_{\Phi} = (D_\mu \Phi)^\dagger (D^\mu \Phi) + m^2_\Phi \Phi^\dagger \Phi - \lambda_{\phi} (\Phi^{\dagger} \Phi)^2 -\lambda_{\Phi H} (\Phi^{\dagger} \Phi) (H^{\dagger} H)
	\label{scalarL}
\end{equation}
where $H$ is the SM Higgs doublet. The scalar fields which acquire non-zero vacuum expectation value (VEV) can be represented as
\begin{align*}
	H = \begin{pmatrix}
		h^+  \\
		\frac{(h+v+i h^I)}{\sqrt{2}} 
	\end{pmatrix}, \;\; \Phi=\frac{\phi+u+ i\phi^I}{\sqrt{2}}.
\end{align*}
The singlet scalar VEV gives rise to $U(1)_X$ gauge boson mass $M_{Z'}=2g'u$ while Higgs doublet gives rise to the usual SM particle masses.

The scalar singlet $\Phi$ also breaks $U(1)_X$ spontaneously down to a remnant $Z_2$ symmetry under which $\Psi_{L,R}$ are 
odd while all other fields are even. As a result, $\Psi_L$ and $\Psi_R$ combine to give a stable DM candidate in the low energy effective theory. The VEV 
of $\Phi$ also generates Majorana masses for fermion DM: $m_L=y_L u/\sqrt{2}$ and $m_R = y_R u/\sqrt{2}$ for $\Psi_L$ and $\Psi_R$ respectively. We 
assume $m_L, m_R \ll M$. As a result, the Dirac fermion $\Psi=\Psi_L + \Psi_R$ splits into two pseudo-Dirac states $\psi_1$ and $\psi_2$ with masses 
$M_1= M-m_+$ and $M_2=M+m_+$, where $m_\pm=(m_L\pm m_R)/2$. The DM Lagrangian after spontaneous symmetry breaking can be written as
\begin{equation}
	\begin{aligned}
		\label{model_Lagrangian}
		\mathcal{L}_{DM} &= \frac{1}{2} \overline{\psi_1} \gamma^\mu \partial_\mu \psi_1 + \frac{1}{2} \overline{\psi_2} \gamma^\mu \partial_\mu \psi_2  -\frac{1}{2}M_1\overline{\psi_1}{\psi_1} -\frac{1}{2}M_2\overline{\psi_2}{\psi_2} + \frac{\epsilon}{2}B^{\alpha \beta}Y_{\alpha\beta}\\  
		& +  i g' Z'_\mu \overline{\psi_1} \gamma^\mu \psi_2 + \frac{1}{2} g' Z'_\mu\frac{m_-}{M}(\overline{\psi_2} \gamma^\mu \gamma^5 \psi_2 - \overline{\psi_1} \gamma^\mu \gamma^5 \psi_1 ) \\
		& + \frac{1}{2}(y_L \cos^2\theta -y_R \sin^2\theta)\overline{\psi_1}\psi_1 \phi
		+ \frac{1}{2}(y_R\cos^2\theta -y_L \sin^2\theta)\overline{\psi_2}\psi_2 \phi
	\end{aligned}
\end{equation}
where $\sin \theta \approx m_-/M$. The mass splitting between the two mass eigenstates is given by $\Delta m =M_2 - M_1= 2m_+ 
= (y_L+y_R) \frac{u}{\sqrt{2}} $. In order to address the XENON1T anomaly, we take $\Delta m \sim 2$ keV. While we stick to such minimal DM models in this work, such Abelian gauge extensions can be motivated from other phenomena like origin of light neutrino masses as well, as discussed in several works including \cite{Adhikari:2008uc, Borah:2012qr, Adhikari:2015woo, Patra:2016shz, Nanda:2017bmi, Barman:2019aku, Biswas:2019ygr, Nanda:2019nqy, Bhattacharya:2020wra, Mahapatra:2020dgk, Borah:2020smw, Okada:2019sbb}.
\section{Dark Matter Self-interaction}
\label{sec3}
As mentioned before, we have an inelastic DM scenario where two DM components with tiny mass splitting of keV scale can populate the universe. We are also considering light mediator of DM interactions which is motivated from SIDM solution to structure formation problems. See \cite{Schutz:2014nka, Blennow:2016gde, Zhang:2016dck} for earlier studies on self-interacting inelastic DM which were primarily motivated from the requirements of avoiding strong direct detection constraints or to explain some anomalous observations at indirect detection experiments like monochromatic photon lines. While in our model DM candidates can interact among themselves either via scalar or vector mediators, we consider only the vector mediator or $Z'$ to be light and hence consider this only to constrain the parameter space from required self-interactions. The relevant Lagrangian for DM interactions with $Z'$ can be rewritten as
\begin{equation}
	\mathcal{L} = i g' Z'_\mu \overline{\psi_1} \gamma^\mu \psi_2 + \frac{1}{2} g' Z'_\mu\frac{m_-}{M}(\overline{\psi_2} \gamma^\mu \gamma^5 \psi_2 - \overline{\psi_1} \gamma^\mu \gamma^5 \psi_1 ).
\end{equation}
Ignoring the second term which is suppressed by $m_{-}/M$, we can write down the corresponding potential for two Majorana fermion DM with a light mediator of dark photon type as \cite{Schutz:2014nka, Blennow:2016gde, Zhang:2016dck, ArkaniHamed:2008qn}
\begin{equation}
	V(r)=\left(
	\begin{array}{cc}
		0 & -\alpha e^{M_Z' r}\\
		-\alpha e^{M_Z' r} &2\Delta m \\
	\end{array}
	\right).\,
\end{equation}
The two body Schrodinger equation for relative motion is
\begin{equation}
	\frac{1}{M_\psi}\nabla^2 \Psi(\vec{r})=\big(V(r) - M_\psi v^2)\Psi(\vec{r})
\end{equation}
where $M_{\psi}$ is the mass of the dark matter, ignoring the tiny mass splitting $\Delta m$ , v is the individual velocity of either of the dark matter particles in the centre of mass frame (half the relative velocity), $\Delta m$ is the mass splitting between two DM candidates and $\Psi(\vec{r})$ is the wave function. Defining dimensionless parameters, $\epsilon_v=\frac{v}{ \alpha}$, $\epsilon_\delta=\sqrt{\frac{2\Delta m}{M_\psi \alpha^2}}$, $\epsilon_{\phi}=\frac{M_{Z'}}{M_\psi \alpha}$ and writing $r\Psi(\vec{r})=\psi(r)$, the s-wave Schrodinger equation is given by
\begin{equation}
	\psi''(r)=\left(
	\begin{array}{cc}
		-\epsilon^2_v & -\frac{e^{\epsilon_Z r}}{r}\\
		-\frac{e^{\epsilon_Z r}}{r} &~~~\epsilon^2_\delta -\epsilon^2_v \\
	\end{array}
	\right)\psi(r)
\end{equation}

\begin{figure}
	\centering
	\includegraphics[scale=0.35]{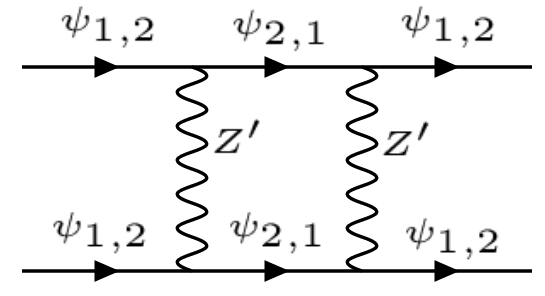}
	\hfil
	\includegraphics[scale=0.5]{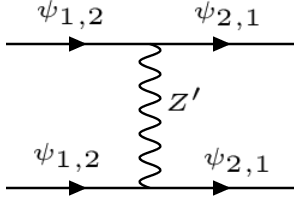}
	\caption{Feynman diagrams for self-interaction.}
	\label{feyn1}
\end{figure}
As shown in Feynman diagrams of figure \ref{feyn1}, DM candidate of one type can scatter off each other while remaining in the same state, only at one-loop level, due to the off-diagonal nature of DM-mediator couplings.  Using these, we constrain the DM parameter space from the required self-interactions at different scales while considering the mass splitting between the two DM candidates to be 2 keV, as favoured from XENON1T excess. The relevant cross sections are given in appendix \ref{appen1}. For a more general analysis, one may refer to \cite{Schutz:2014nka}.

\begin{figure}
	$$
	\includegraphics[scale=0.35]{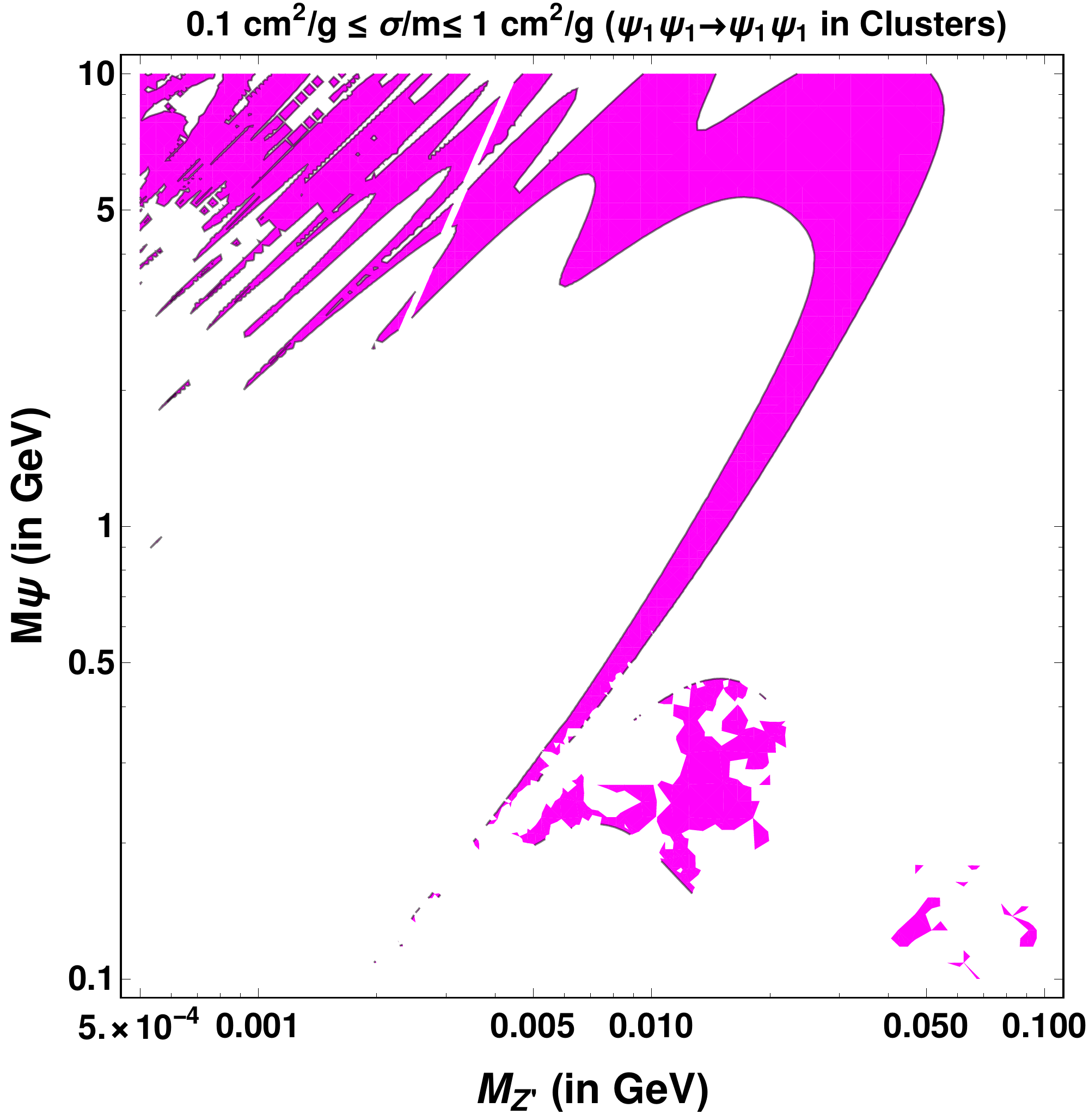} 
	\includegraphics[scale=0.35]{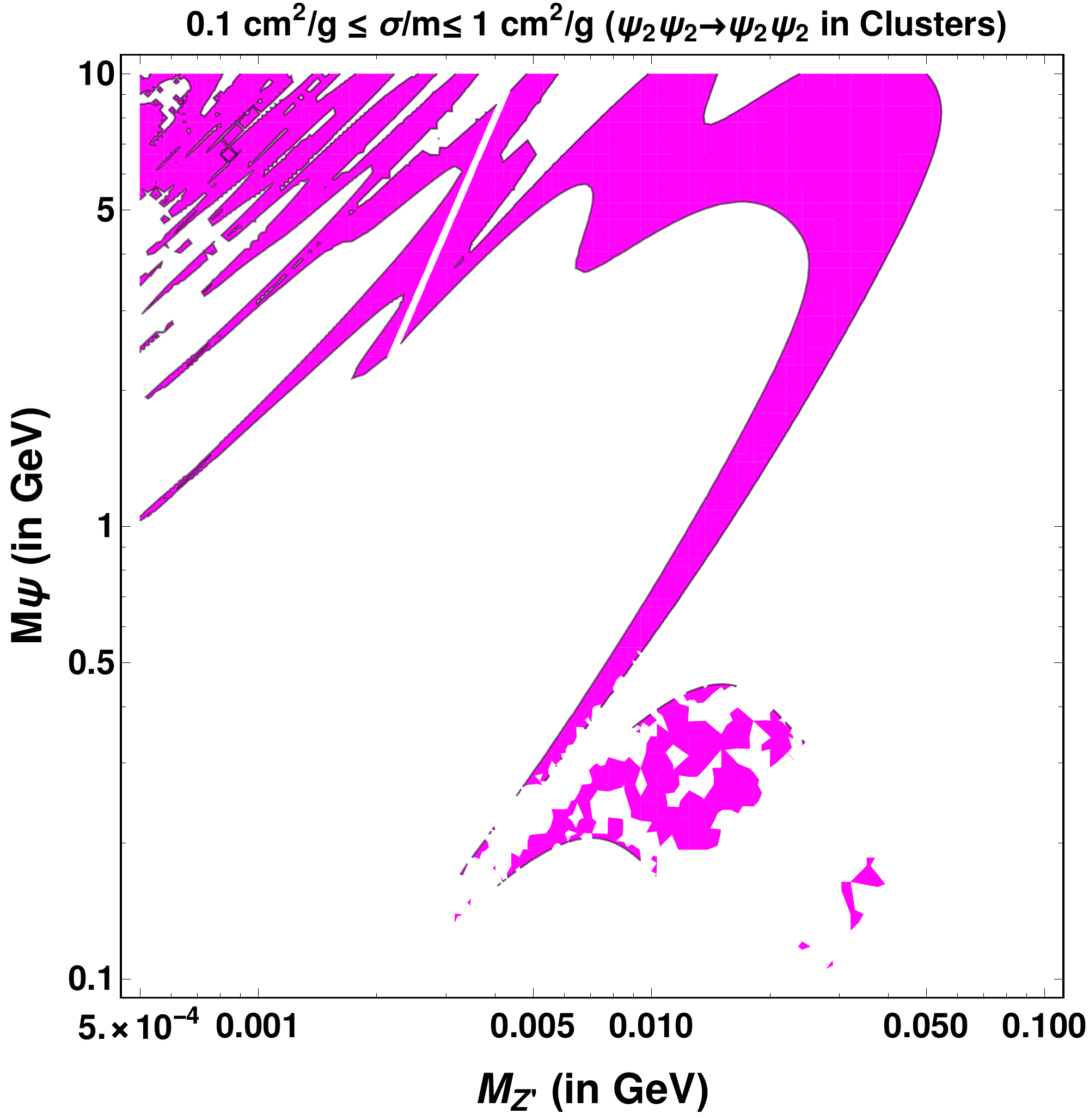}
	$$
	$$
	\includegraphics[scale=0.35]{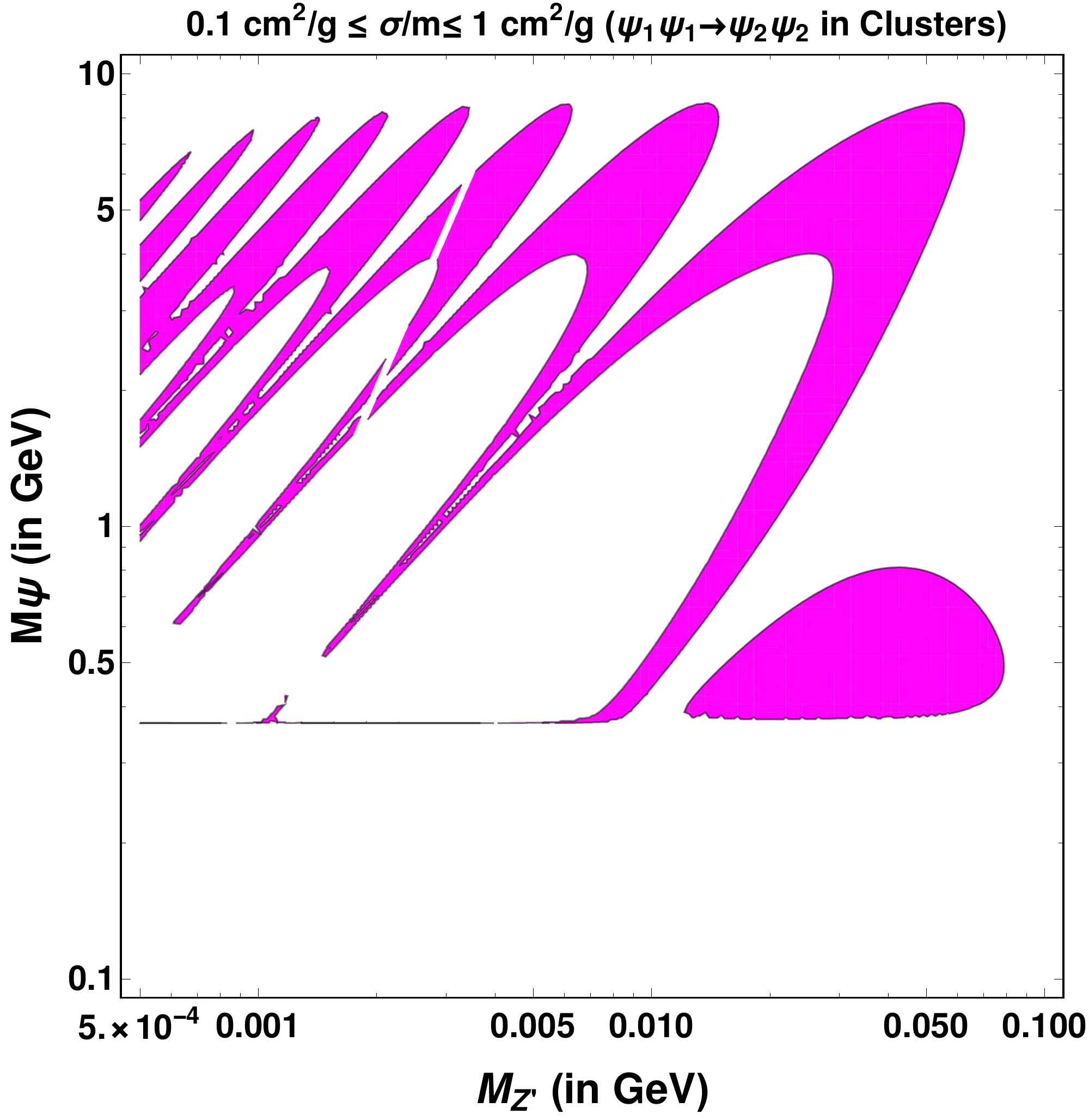} 
	\includegraphics[scale=0.35]{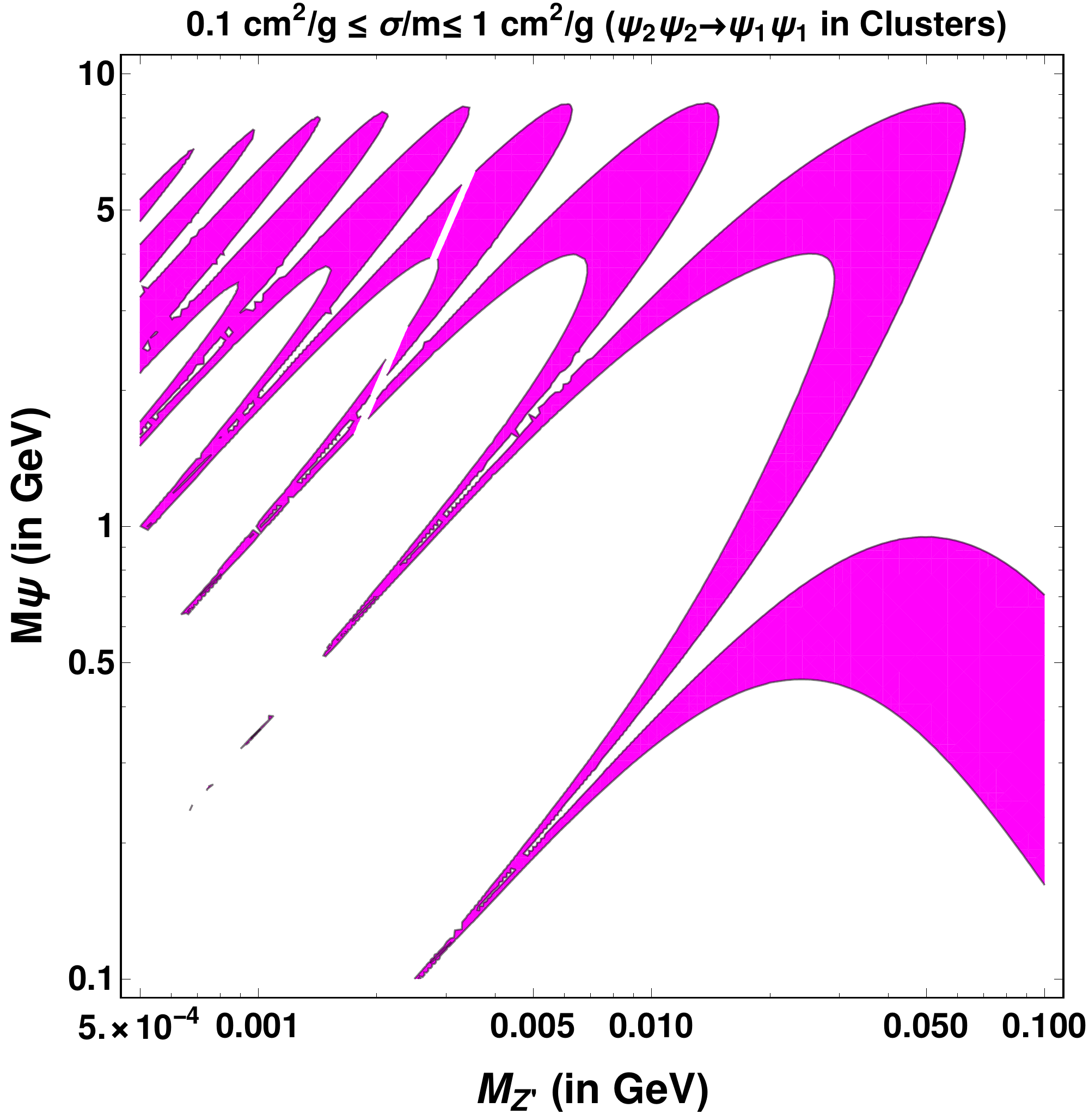}
	$$ 
	\caption{\footnotesize{Self-interaction cross-section ($\sigma/m$) in the range $0.1-1 {\rm cm}^2/{\rm g}$ (light pink coloured region) for clusters ($v\sim1000 km/s$). Top left (right) panel: elastic scattering of ground (excited) to ground (excited) state. Bottom left (right) panel: up (down) scattering of ground (excited) to excited (ground) state.}}
	\label{sidmfig1}
\end{figure}
\begin{figure}
	$$
	\includegraphics[scale=0.35]{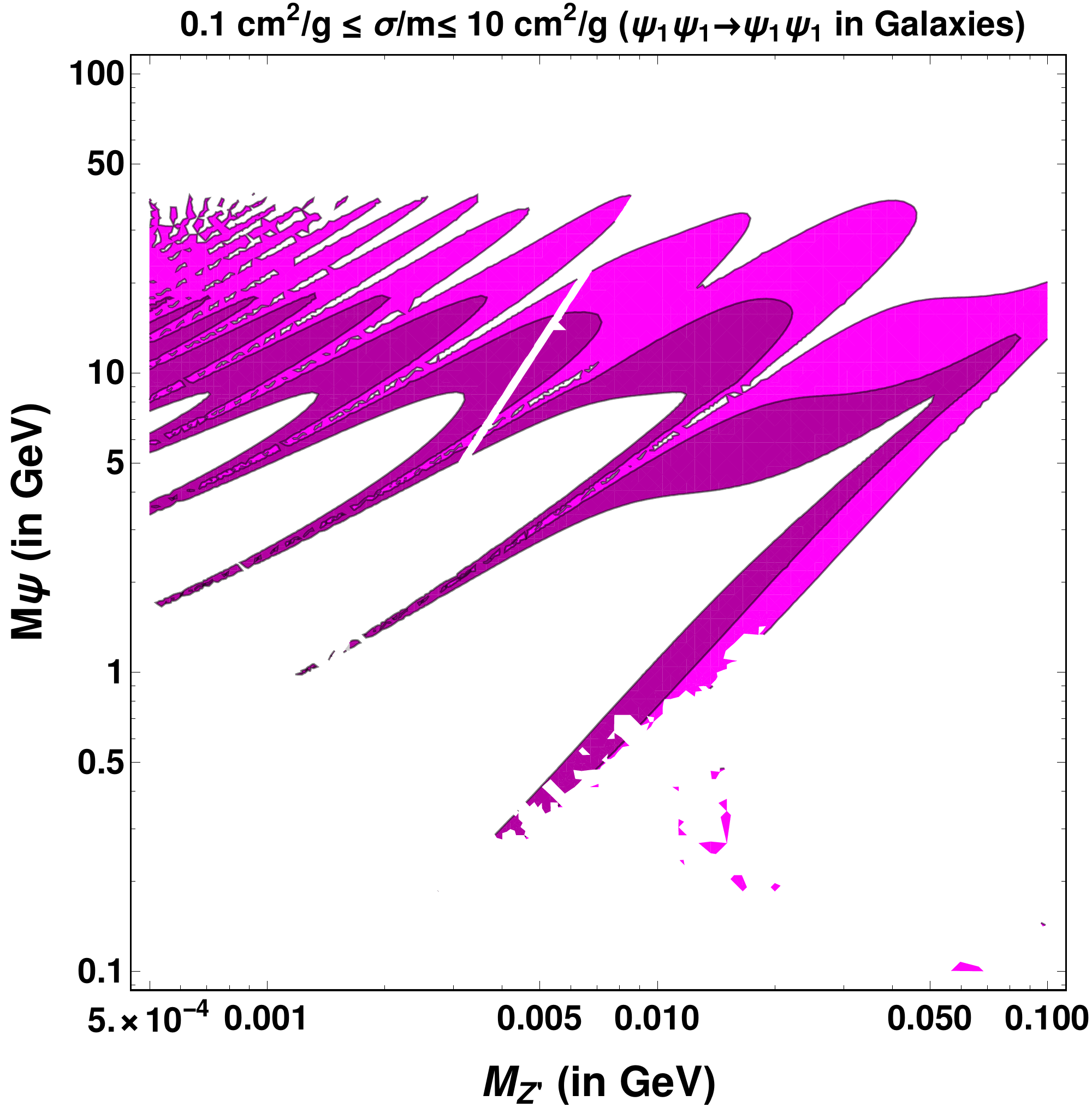} 
	\includegraphics[scale=0.35]{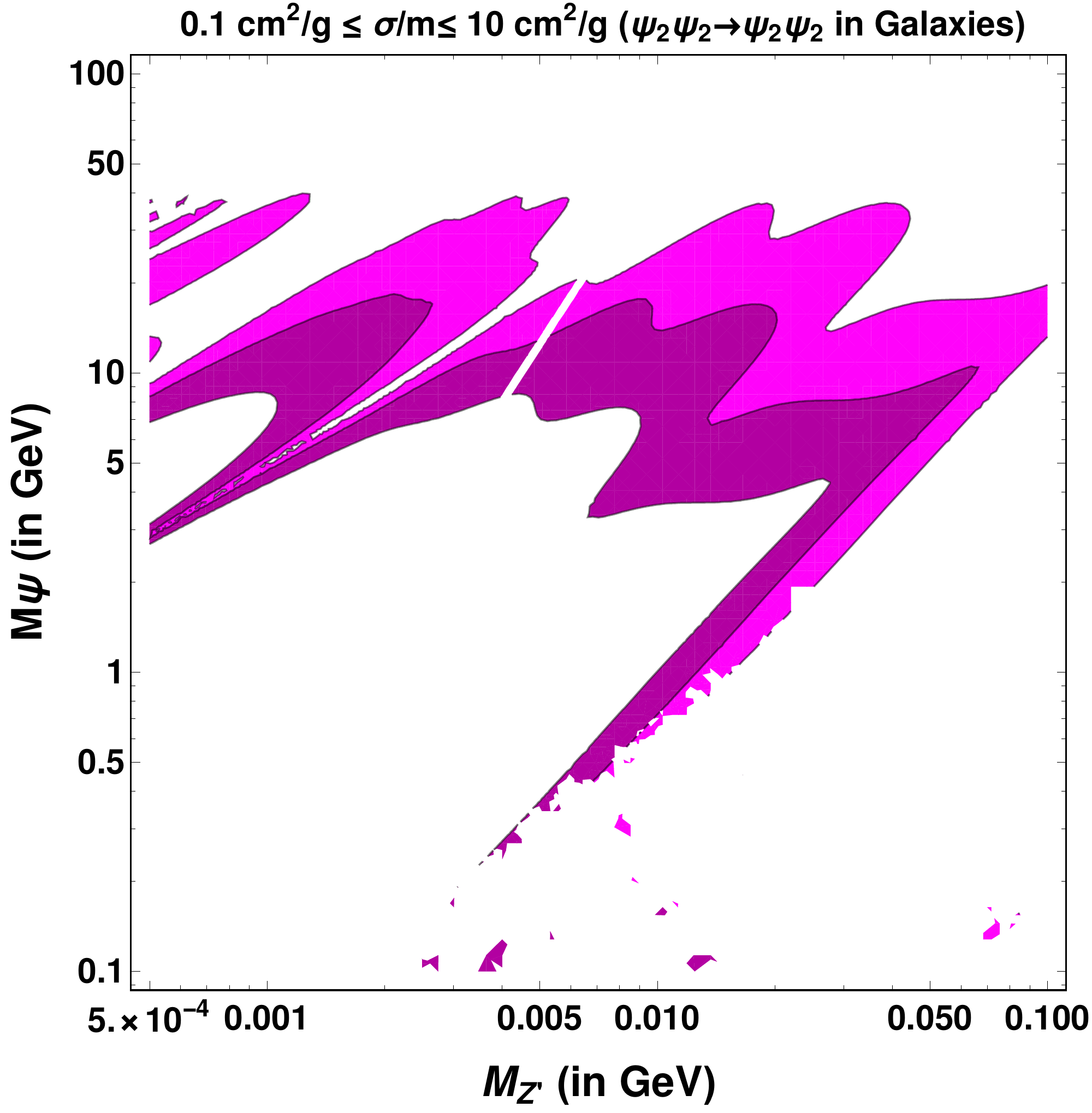}
	$$
	$$
	\includegraphics[scale=0.35]{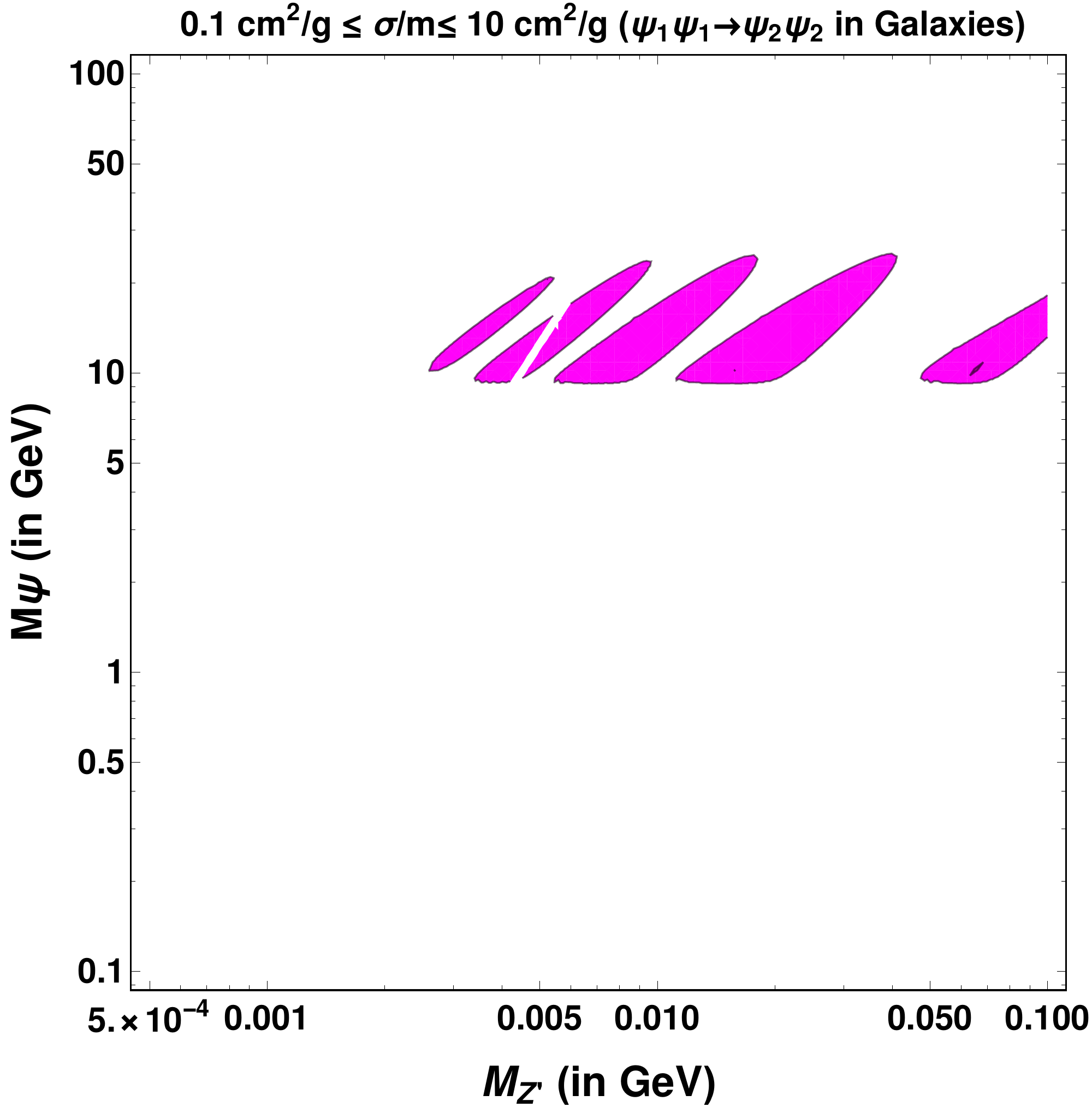} 
	\includegraphics[scale=0.35]{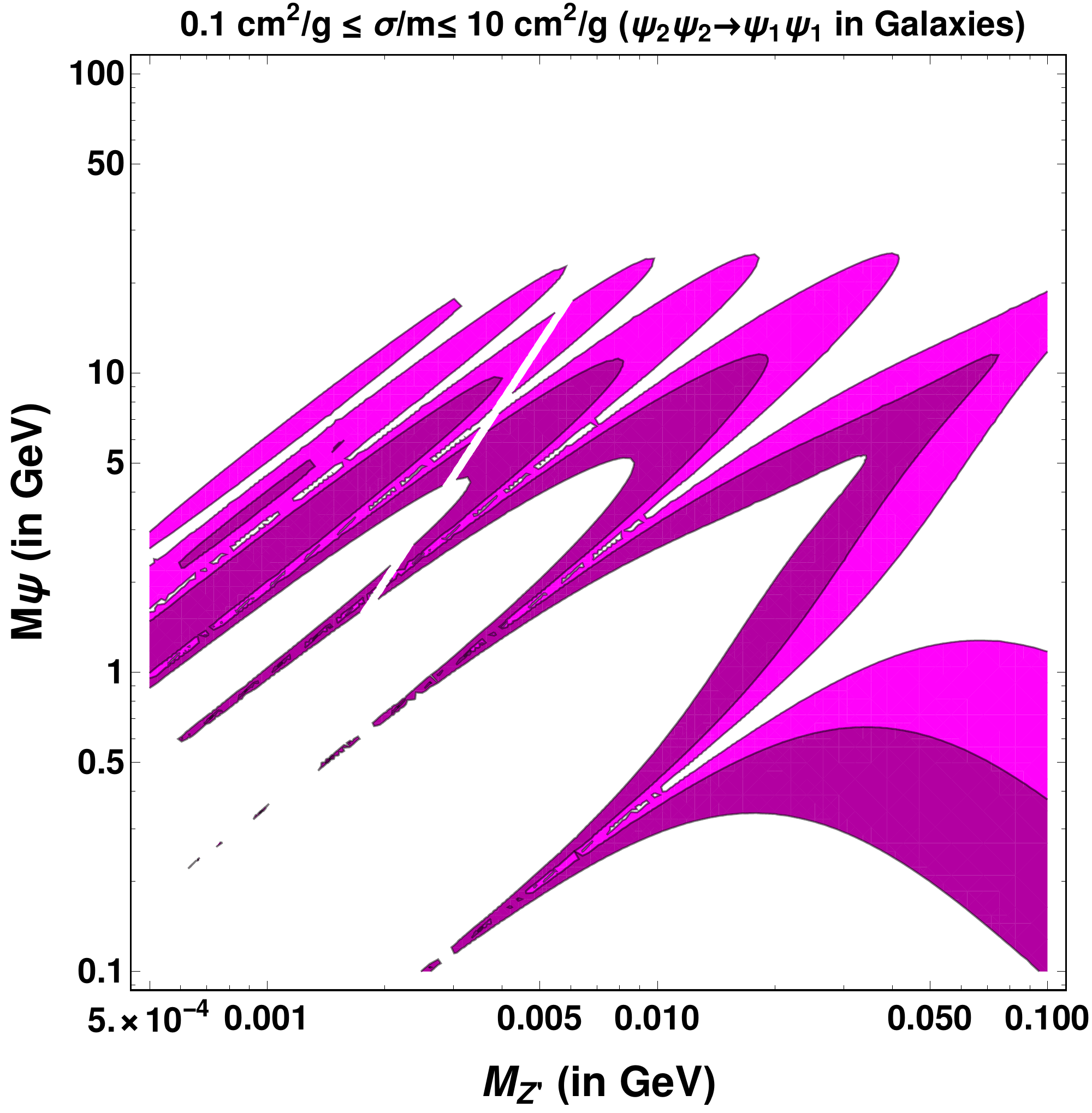}
	$$ 
	\caption{\footnotesize{Self-interaction cross-section ($\sigma/m$) in the range $0.1-10 \;{\rm cm}^2/{\rm g}$ for galaxies ($v\sim200 \; {\rm km/s}$). Light pink coloured region represents the parameter space where $0.1 \;{\rm cm}^2/{\rm g} < \sigma/m <1 \; {\rm cm}^2/{\rm g}$, dark pink colour represents regions of parameter space where $1 \; {\rm cm}^2/{\rm g} < \sigma/m < 10 \; {\rm cm}^2/{\rm g}$. Top left (right) panel: elastic scattering of ground (excited) to ground (excited) state. Bottom left (right) panel: up (down) scattering of ground (excited) to excited (ground) state.}}
	\label{sidmfig2}
\end{figure}
\begin{figure}[ht]
	$$
	\includegraphics[height=5cm]{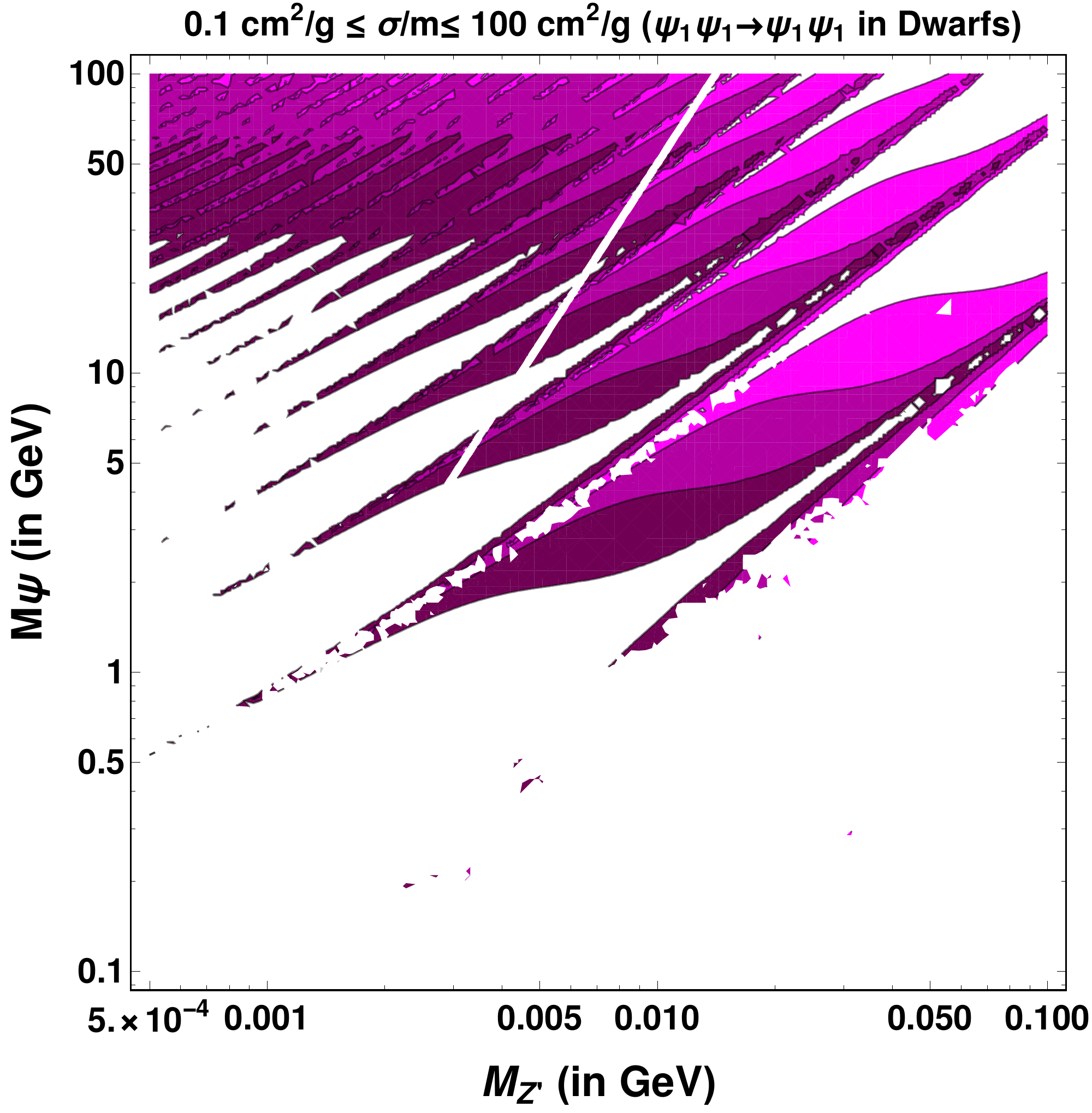} 
	\includegraphics[height=5cm]{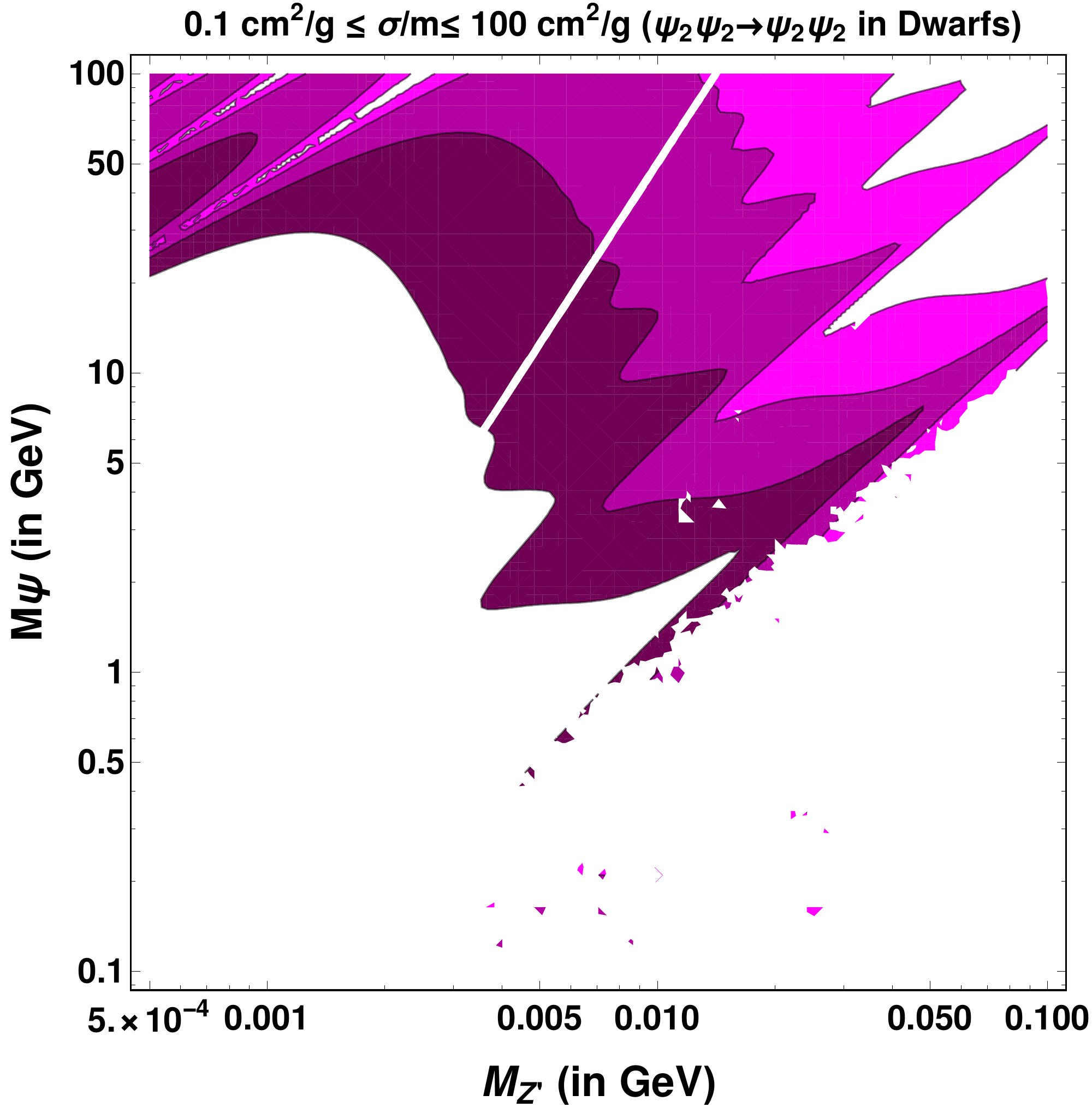}
	\includegraphics[height=5cm]{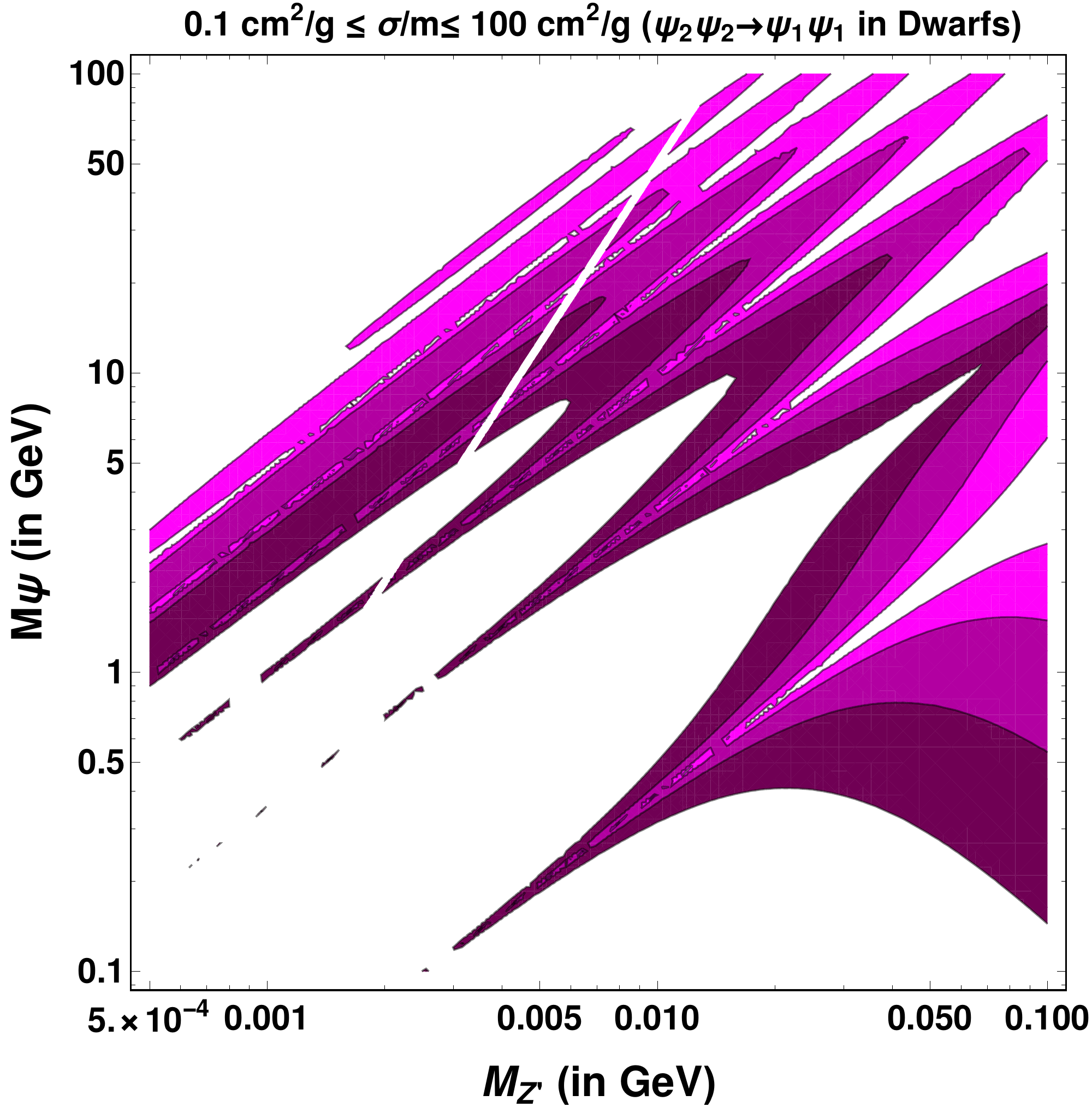}
	$$ 
	\caption{\footnotesize{Self-interaction cross-section in the range $0.1-100 \; {\rm cm}^2/{\rm g}$ for dwarfs ($v\sim10 \; {\rm km/s}$). Light pink colour represents regions of parameter space where $0.1 \; {\rm cm}^2/{\rm g} < \sigma/m <1 \;{\rm cm}^2/{\rm g}$; dark pink colour represents regions of parameter space where $1 \; {\rm cm}^2/{\rm g} < \sigma/m <10 \; {\rm cm}^2/{\rm g}$; maroon colour  represents regions of parameter space where $10 \; {\rm cm}^2/{\rm g} < \sigma/m < 100 \; {\rm cm}^2/{\rm g}$. Left (middle) panel: elastic scattering of ground (excited) to ground (excited) state. Right panel: down scattering of excited to ground state.}}
	\label{sidmfig3}
\end{figure}

Using these self-interaction cross sections and using the required $\sigma/m$ from astrophysical observations at different scales, we constrain the parameter space of the model in terms of DM $(\psi_{1,2})$ and mediator $Z'$ masses. As our study is motivated from explaining the XENON1T excess, we keep the required mass splitting between two DM candidates to be 2 keV. In figure~\ref{sidmfig1}, we show the allowed parameter space in DM mass versus $Z'$ mass plane which gives rise to the required DM self-interaction cross-section ($\sigma/m$) in the range $0.1-1~{\rm cm}^2/{\rm g}$ for clusters ($v\sim1000~ \rm km/s$). The corresponding region of parameter space for galaxies ($v\sim 200~ \rm km/s$)and dwarf galaxies ($v\sim 10~ \rm km/s$) are shown in figure~\ref{sidmfig2} and figure~\ref{sidmfig3} respectively. It should be noted that for dwarf galaxies, due to smaller DM velocities we do not get sufficient self-interaction cross section ($\sigma/m$) from up scattering processes in the entire parameter space considered and hence the corresponding plot is not shown in figure~\ref{sidmfig3}. This is due to the fact that, lighter DM, due to low velocities, do not have sufficient kinetic energies to scatter efficiently into heavier DM resulting in a large self-interaction cross section. We will finally compare these regions of parameter space of GeV scale DM mass in the context of XENON1T excess and other phenomenological constraints.

\begin{figure}
	$$
	\includegraphics[scale=0.4]{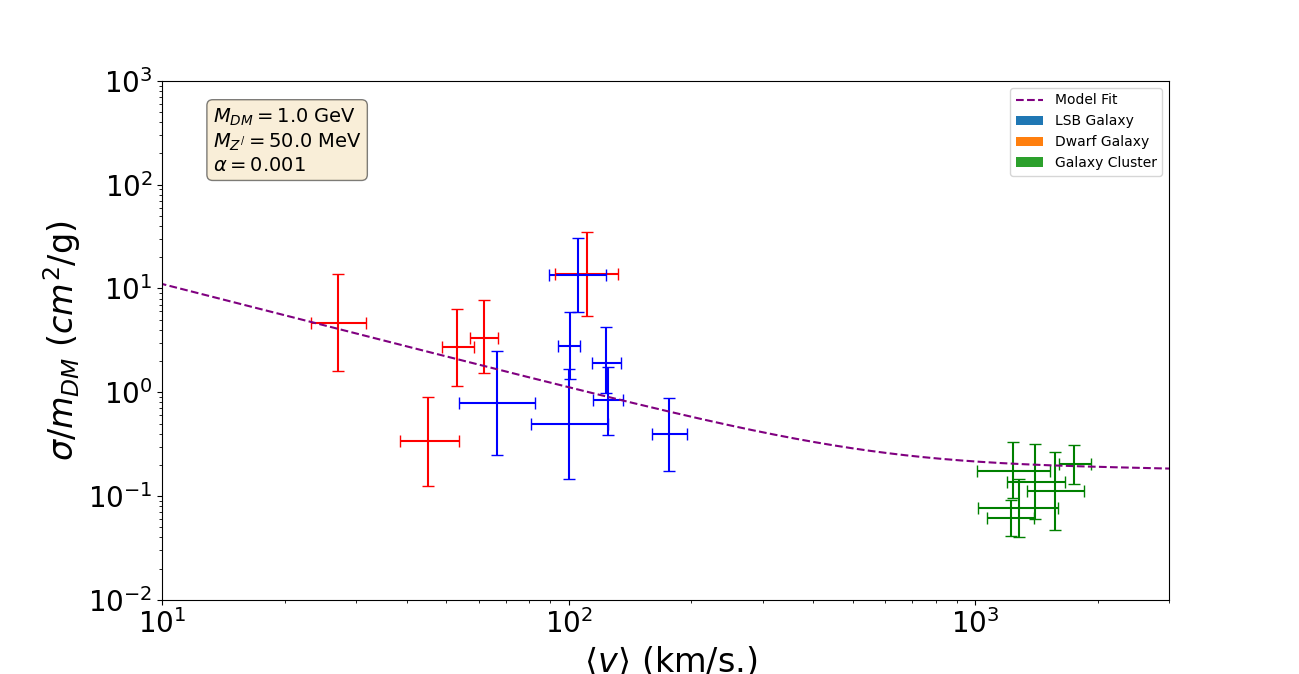}
	$$
	\caption{\footnotesize{The self-interaction cross section per unit mass of DM as a function of average collision velocity.}}
	\label{astrofit}
\end{figure}

The self-interaction cross section per unit mass of DM as a function of average collision velocity is shown in figure~\ref{astrofit} as measured from astrophysical data. The data includes measurements from dwarfs (red), LSBs (blue) and clusters (green)~\cite{Kaplinghat:2015aga,Kamada:2020buc}. The purple dashed curve corresponds to the velocity-dependent cross section from our model for a particular set of benchmark values (i.e $M_{\psi}=1~\rm GeV$, $M_{Z'}=50~\rm MeV$ and $\alpha=0.001$) allowed from all relevant phenomenological constraints. It is clear from the figure that the model proposed here can explain the astrophysical observation of self-interaction of DM appreciably well. See \cite{Alvarez:2019nwt} for discussions on astrophysical probes of such inelastic DM with a light mediator.

\section{Dark Matter Production}
\label{dmrelic}
\begin{figure}[ht]
	$$
	\includegraphics[scale=0.6]{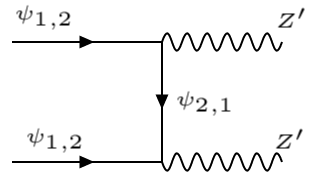} 
	\includegraphics[scale=0.6]{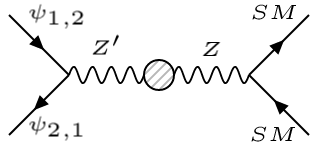}
	$$ 
	\caption{\footnotesize{Feynman diagrams for dominant number changing processes of DM.}}
	\label{thermal_relic}
\end{figure}

\begin{figure}
	\centering
	\includegraphics[scale=0.5]{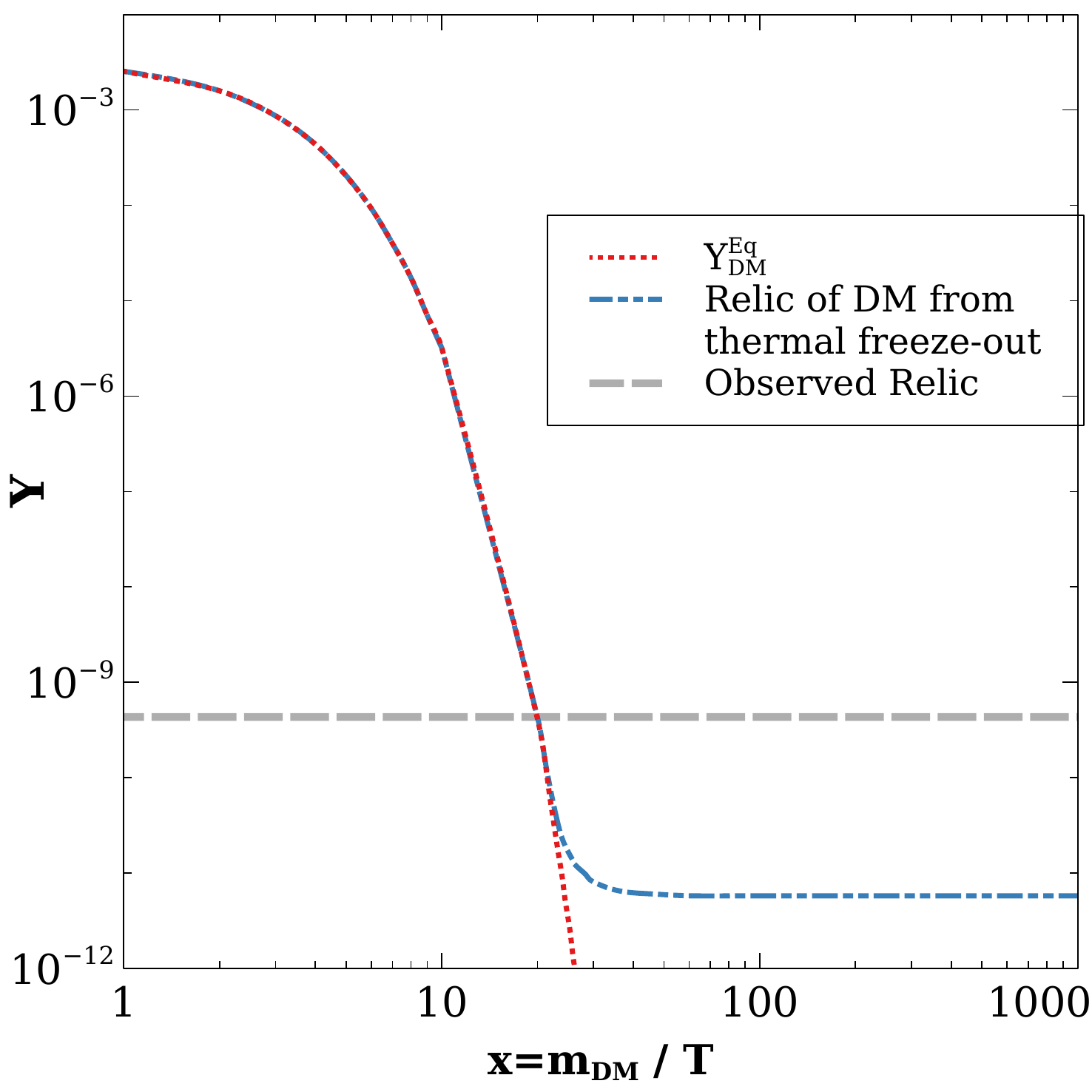}
	\caption{\footnotesize{Relic abundance of DM assuming it to be produced thermally in the early universe followed by thermal freeze-out. The thermal relic is under-abundant by two orders of magnitudes.}}
	\label{Thermal_relics}
\end{figure}

While several production regimes for self-interacting DM exist in the literature, we first consider the usual $2 \leftrightarrow 2$ vector portal interactions. 
While DM can interact with itself via $Z'$  as well as singlet scalar interactions, we consider the vector portal to be dominant due to light $Z'$. On the other 
hand, DM can interact with the SM bath only via kinetic mixing of neutral vector bosons. These dominant number changing processes are shown in 
figure~\ref{thermal_relic}. While DM-SM interactions via kinetic mixing is responsible for production of DM from the thermal bath, the dark sector interactions 
can be important to decide final abundance of DM. Since from SIDM point of view we consider heavier DM mass compared to the mediator $m_{\psi_{1,2}} > m_{Z'}$, DM can have a large annihilation cross section to $Z'$ affecting its relic abundance. For example, the thermal averaged cross section for the t-channel process $\psi_1 \psi_1 \rightarrow Z' Z'$ shown in the left panel of figure~\ref{thermal_relic} is
\begin{equation}
	\langle\sigma v\rangle \sim \frac{\pi \alpha^2_x}{M^2_\psi}
\end{equation}
where $\alpha_x=g'^2/(4\pi) $ and for typical gauge coupling and DM mass of our interest we have $\alpha_x \sim 0.001, M_\psi \sim 1$ GeV. This leads to a cross section which is at least two order of magnitudes larger compared to the typical annihilation cross section of thermal DM. This reduces the relic abundance by same order of magnitudes, as seen from figure \ref{Thermal_relics} showing the comoving number density of DM, assuming it to be a purely thermal relic. Before calculating DM relic, we first compare rates of different annihilation processes. Note that for the purpose of numerical analysis, the model has been implemented in \texttt{LanHEP} \cite{Semenov:2014rea} and \texttt{CalcHEP} \cite{Belyaev:2012qa} and the cross-sections required has been fed into \texttt{Mathematica} \cite{Mathematica} from \texttt{CalcHEP}.

\begin{figure}
	\centering
	\includegraphics[scale=0.6]{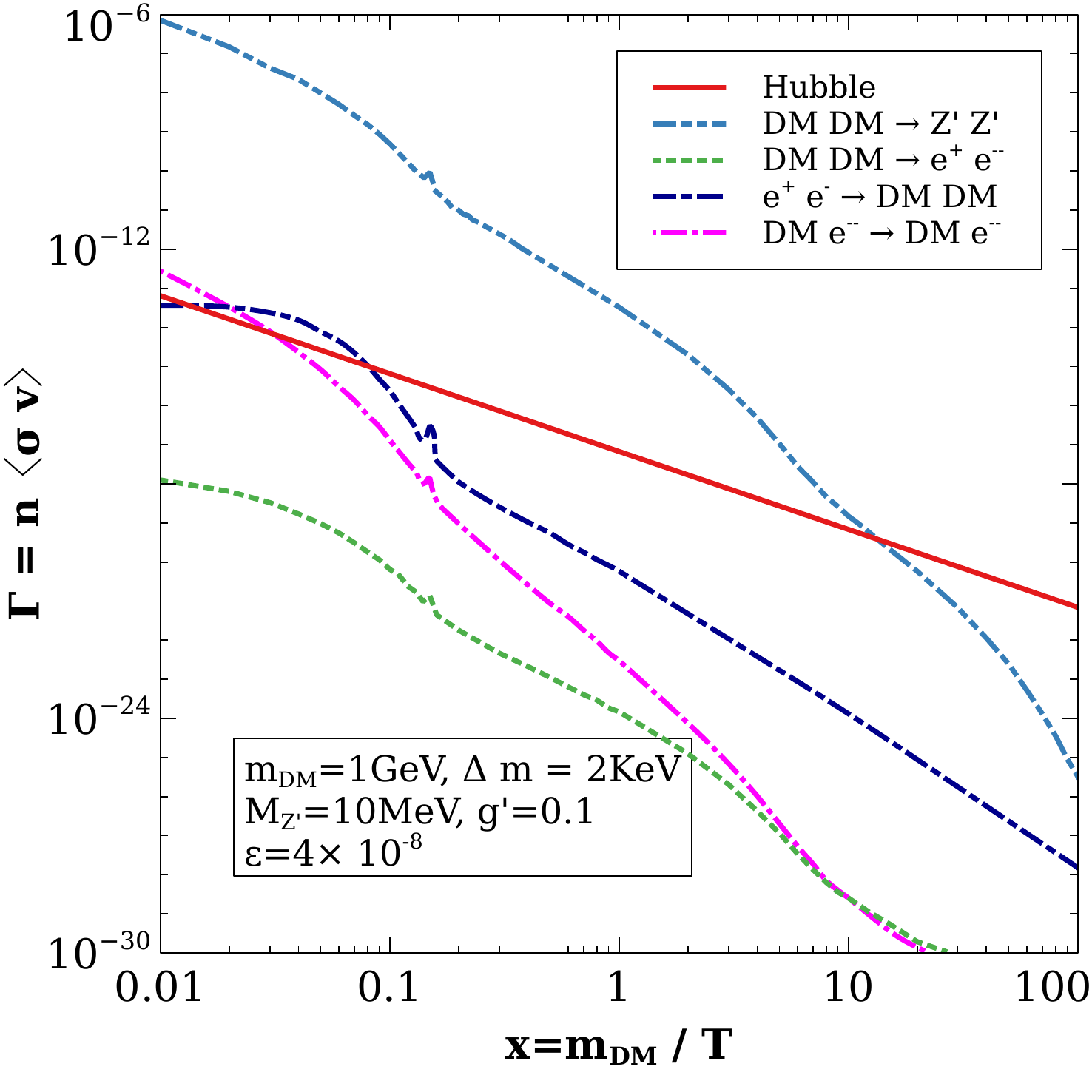}
	\caption{\footnotesize{Decoupling of different processes from the thermal plasma.}}
	\label{Decoupling}
\end{figure}

Although dark sector interaction rates are large as mentioned above, the DM-SM interactions are suppressed due to tiny kinetic mixing chosen to 
realise the required XENON1T excess. We check the relevant DM-SM processes and find that for the chosen sub-GeV regime and kinetic mixing, DM 
never attains chemical equilibrium with the SM bath. While relevant cross sections are given in appendix \ref{appen2}, we compare different 
interaction rates in figure~\ref{Decoupling}. In calculating the interaction rates we consider the light SM degrees of freedom to be in 
equilibrium, while the DM number density is calculated by solving the appropriate Boltzmann equation considering freeze-in production of 
DM \cite{Hall:2009bx} from SM bath. This happens dominantly from $2 \rightarrow 2$ processes where SM fermions in equilibrium at GeV temperatures 
can contribute to the production of DM. Since the production happens from the thermal bath, it saturates at a temperature similar to that of 
DM mass. On the other hand, DM produced this way keeps annihilating into $Z'$ bosons due to large self-interactions further diluting the 
DM abundance. Clearly, almost all the $2 \rightarrow 2$ processes remain out of equilibrium as the corresponding interaction rates remain 
below Hubble expansion rate seen from figure~\ref{Decoupling}. Only DM annihilation rate into $Z'$ boson remains in equilibrium for a longer 
epoch while DM-SM kinetic equilibrium is reached for a very short epoch. We also check that the freeze-in production of DM from thermal bath, 
followed by dark sector freeze-out is insufficient to produce the correct DM relic for the region of our interest. This is due to the large 
annihilation rates of DM into $Z'$ bosons keeping DM under-abundant after dark sector freeze-out. This requires an additional source of DM 
relic which we consider to be a SM singlet scalar $\eta$. The singlet scalar freezes out in the early Universe via the process: $\eta^\dagger \eta \to 
H^\dagger H$, and decays back to DM after the dark sector freezes-out, thus filling the deficit in relic abundance. The relevant Lagrangian 
for $\eta$ decay is given by:
\begin{equation}
	\mathcal{L}=
	\frac{1}{2}\lambda_1\overline{\psi_1}\psi_1 \eta
	+\frac{1}{2}\lambda_2\overline{\psi_2}\psi_2 \eta.
\end{equation}
If the thermal averaged annihilation cross-section of $\eta^\dagger \eta \to H^\dagger H$ is smaller than the usual freeze-out cross-section 
of a WIMP, {\it i.e.}, $\langle \sigma|v|\rangle_F = 3 \times 10^{-26} {\rm cm}^3/s$, then the relic of $\eta$ can be larger than the observed 
DM abundance. As a result the late decay of $\eta \to \psi_i\psi_j$ can give rise to ample amount of DM. In Eq. \ref{boltzmann}, we use appropriate 
Boltzmann equations to get the correct relic density of DM. While we incorporate this additional scalar singlet $\eta$ only to satisfy DM relic through its late decay, it can serve other purposes as well. One such possibility is to realise cosmic inflation. There have been proposals where a single scalar field is shown to play the role of inflation as well as thermal DM whose relic is generated via usual freeze-out. For example, see \cite{Borah:2018rca} and references therein. The same idea can be implemented here as well while noting that the scalar field is not perfectly stable but decays at late epochs into the DM. We however, do not discuss such additional roles the scalar singlet might play in our minimal scenario discussed here.

From figure~\ref{Decoupling}, it is evident that the process $DM e \rightarrow DM e$ which is responsible for keeping both the dark and visible sector in kinetic equilibrium decouples around $x \sim 0.03$, after which the temperature of the dark sector (denoted by $T'$) evolves independently of the thermal bath (temperature T) until $x \sim 100$ when all the dark sector particles becomes non-relativistic (and hence ceases to contribute to the relativistic degrees of freedom). Between these two epochs, the ratio of the two temperatures can be obtained by conserving the total entropy separately in the two sectors. Considering the kinetic decoupling temperature to be $T_D$, we can relate the temperature of the two sectors as 
\begin{equation}
	\frac{T'}{T} = \left( \frac{g^{\rm SM}_{*s} (T)}{g^{\rm SM}_{*s} (T_D)} \right)^{1/3}.
	\label{eqdec}
\end{equation}
Here $g^{\rm SM}_{*s} (T)$ is the relativistic entropy degrees of freedom in the standard model which goes into the calculation of relativistic entropy density $s(T)=\frac{2\pi^2}{45}g_{*s}(T)T^3$. Since the above relation \eqref{eqdec} is for $T<T_D$, we naturally have $g^{\rm SM}_{*s} (T) < g^{\rm SM}_{*s} (T_D)$ leading to $T'< T$. This is also understood from the fact that SM bath temperature receives additional entropy contributions from the species which keep getting decoupled gradually. Within the decoupled dark sector itself, the DM particles can transfer their entropies into lighter $Z'$ bosons once $T'$ falls below DM mass. This corresponds to an enhancement of dark sector temperature for $T'< m_{\rm DM}$ by $(13/6)^{1/3}$, a factor close to unity. We have ignored this additional enhancement in the calculations.

Due to different temperatures of dark sector and SM bath after some epoch, we accordingly divide the range of integration for solving the Boltzmann equations as follows:
\begin{itemize}
	\item  From the epoch of reaching kinetic equilibrium between DM-SM sectors till $x < 0.03$ (see figure~\ref{Decoupling}), both the dark and the visible sectors share the same temperature $T=T'$.
	\item One with $0.03<x<100$ where the dark sector is decoupled from the thermal bath and its temperature evolves according to \eqref{eqdec}. 
\end{itemize}
Accordingly, one can define a new dimensionless parameter and relate to the usual parameter $x=\frac{m_{\rm DM}}{T}$ as
\begin{equation}
	x'=\frac{m_{\rm DM}}{T'}=\Big(\frac{T}{T'}\Big) x
\end{equation}
We can now write down the Boltzmann equations for two DM candidates $\psi_{1,2}$ and the scalar singlet $\eta$ whose late decays into DM is crucial to generate correct DM relic. Unlike DM whose interactions with the SM bath are suppressed due to tiny kinetic mixing, the scalar singlet can be in thermal equilibrium with the SM due to large quartic couplings followed by freeze-out\footnote{This is, to some extent, similar to the super-WIMP dark matter formalism \cite{Feng:2003uy}.}. Thus, we define comoving number densities of these particles as $Y_{\psi_{1,2}}=n_{\psi_{1,2}}/s'(T'(T)), Y_{\eta} = n_{\eta}/s(T)$. The relevant coupled Boltzmann equations can then be written as
\begin{equation}\label{boltzmann}
	\begin{aligned}
		\frac{dY_\eta}{dx'}&=-\frac{s(M_\psi)}{x'^2 H(M_\psi)\Big( \frac{T'}{T}\Big)}\langle \sigma v\rangle_{\eta \eta \to H H} (Y^2_{\eta}-(Y^{eq}_{\eta})^2) -\frac{x' \Big( \frac{T'}{T}\Big)^2(\langle \Gamma_{\eta\rightarrow \overline{\psi_1}\psi_1}\rangle + \langle\Gamma_{\eta\rightarrow \overline{\psi_2}\psi_2 }\rangle)}{H(M_\psi)} Y_\eta;\\
		&\frac{dY_{\psi_1}}{dx'}=\Big(\frac{T'}{T}\Big)^2\Bigg[\frac{s(M_\psi)}{x'^2 H(M_\psi)}\Big(\frac{g'_{*s}(T_D)}{g_{*s}(T_D)} \Big) \Big( \langle \sigma v\rangle_{e^+e^- \to \psi_1 \psi_1} (Y^{eq}_{\psi_1})^2 - \langle \sigma v\rangle_{ \psi_1 \psi_1 \to Z'Z'} Y^2_{\psi_1}
		\\&+ \langle \sigma v\rangle_{ \psi_2 \psi_2 \to \psi_1 \psi_1} \big(Y^2_{\psi_2}-\frac{ (Y^{eq}_{\psi_2})^2}{ (Y^{eq}_{\psi_1})^2}Y^2_{\psi_1}\big)\Big)+ \frac{x'\Big(\frac{g_{*s}(T_D)}{g'_{*s}(T_D)}\Big) \langle \Gamma_{\eta\rightarrow \overline{\psi_1}\psi_1}\rangle}{H(M_\psi)} Y_\eta\Bigg]; \\
		&\frac{dY_{\psi_2}}{dx'}=\Big(\frac{T'}{T}\Big)^2\Bigg[\frac{s(M_\psi)}{x'^2 H(M_\psi)}\Big(\frac{g'_{*s}(T_D)}{g_{*s}(T_D)} \Big) \Big( \langle \sigma v\rangle_{e^+e^- \to \psi_2 \psi_2} (Y^{eq}_{\psi_2})^2 - \langle \sigma v\rangle_{ \psi_2 \psi_2 \to Z'Z'} Y^2_{\psi_2}
		\\&-\langle \sigma v\rangle_{ \psi_2 \psi_2 \to \psi_1 \psi_1} \big(Y^2_{\psi_2}-\frac{ (Y^{eq}_{\psi_2})^2}{ (Y^{eq}_{\psi_1})^2}Y^2_{\psi_1}\big) \Big) + \frac{x' \Big(\frac{g_{*s}(T_D)}{g'_{*s}(T_D)}\Big)\langle \Gamma_{\eta\rightarrow \overline{\psi_2}\psi_2}\rangle}{H(M_\psi)} Y_\eta\Bigg]
	\end{aligned}
\end{equation}
where,  $x'=\frac{m_{\rm DM}}{T'}=\frac{M_{\psi}}{T'}$, $s(M_\psi)=\frac{2\pi^2}{45}g_{*s}M^3_{\psi}$ and $H(M_\psi)=1.67 g^{1/2}_*\frac{M^2_\psi}{M_{Pl}}$. Here $M_{\psi} \approx M_1 \approx M_2$, ignoring the tiny mass splitting $\Delta m$.

\begin{figure}[h]
	\centering
	\includegraphics[scale=0.6]{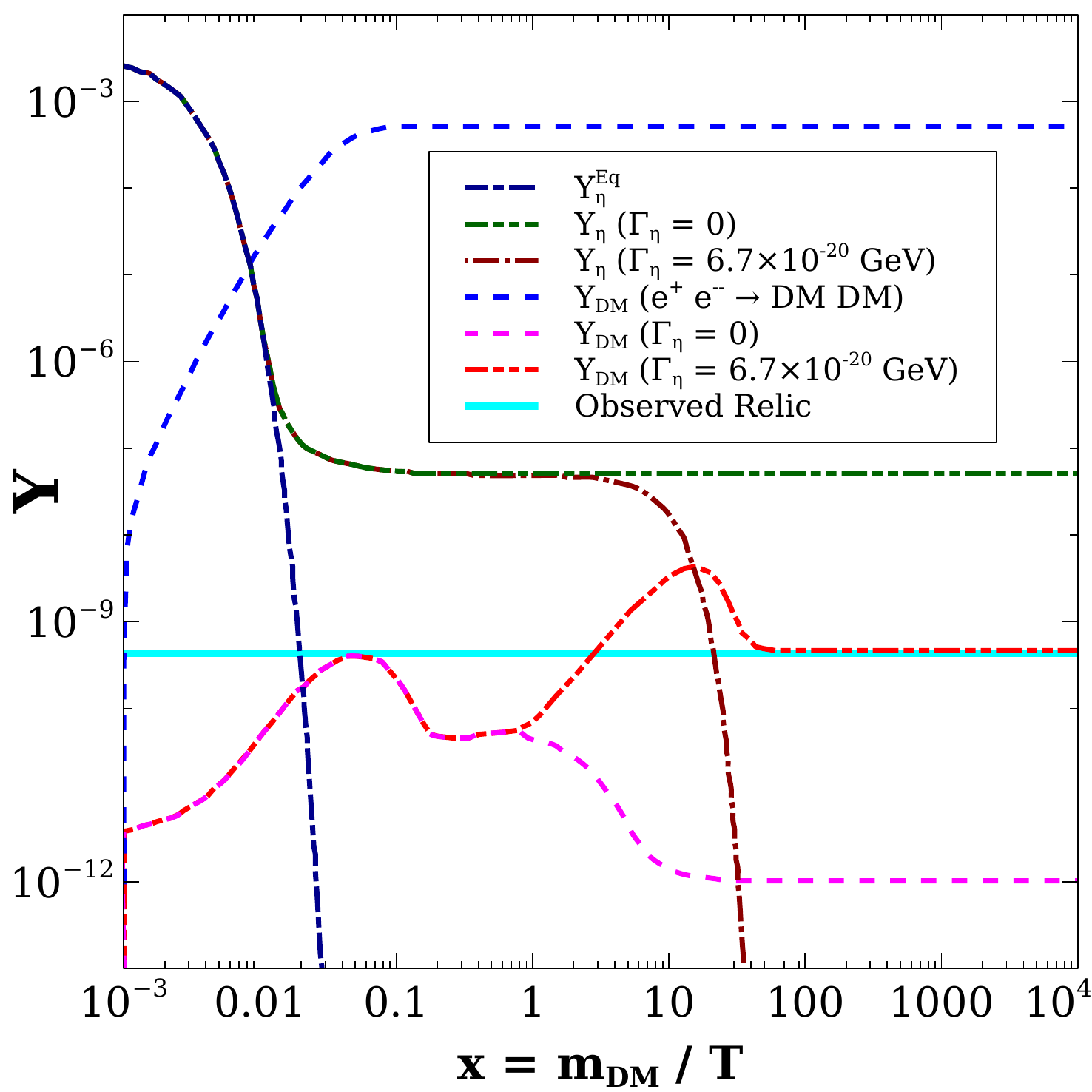}
	\caption{\footnotesize{Comoving number densities of DM and scalar singlet for different cases.}}
	\label{relic}
\end{figure}

We solve these coupled Boltzmann equations taking into account of different temperatures of DM and SM sectors after kinetic decoupling, as given in \eqref{eqdec}. The corresponding evolutions of different comoving number densities are shown in figure~\ref{relic}. In figure~\ref{relic}, the dot-dashed dark blue line shows the equilibrium number density of the singlet scalar $\eta$ with mass $m_\eta \sim 1$ TeV, which was initially in thermal equilibrium with the SM bath. As its interaction rates falls below the expansion rate, it freezes out leaving a thermal relic, shown by the green dot-dashed line, assuming it to be stable. The blue dot-dashed line shows the freeze-in production of DM only from the process $e^+e^- \rightarrow {\rm DM \;DM}$ without considering subsequent annihilation of DM into $Z'$ pairs.  When we take into account both its production from $e^+e^- \rightarrow DM DM$ and subsequent annihilations into $Z'$ bosons via ${\rm DM \; DM} \rightarrow Z'Z'$ its abundance is depicted by the pink line. The sharp contrast is due to the strong ${\rm DM \; DM} \rightarrow Z'Z'$ annihilation rate which reduces the abundance of DM produced from freeze-in. As the number density of DM increases due to freeze-in production, the annihilation rate into $Z'$ pairs also increases leading to the first depletion in the pink line around $x=0.1$. Shortly after that, DM production from freeze-in again balances DM annihilation rate leading to a plateau region all the way till $x=1$. However, since freeze-in production from thermal bath becomes negligible beyond $x=1$, we see further depletion in DM density due to its annihilation into $Z'$ pairs leaving an under-abundant relic beyond $x=10$. Note that, at this point we have not considered scalar decay contribution to DM.

Since freeze-in production of DM from the thermal bath followed by DM annihilation into $Z'$ pairs lead to under-abundant relic density, we now consider the additional contribution from scalar singlet decay. The red dot-dashed line shows the evolution of comoving number density of DM after taking scalar decay contribution into account. The corresponding evolution of the scalar number density is shown by the maroon coloured dot-dashed line. Clearly, once the number density of the scalar falls due to its decay, the DM number density gets uplifted. Once the decay is complete, DM relic also saturates beyond $x \approx 30$. It should be noted that, the 
scalar decay occurs after DM annihilation to $Z'$ pairs freezes out around $x =10$ to avoid further depletion. Also, while considering freeze-in production of DM from the thermal bath, we considered the contribution of electron-positrons only, for simplicity. If we consider all the particles in the thermal bath, we will get more freeze-in production of DM and the final required abundance of DM can be realised by appropriate tuning of scalar decay width without affecting rest of the analysis related to self-interaction and XENON1T excess.

\begin{figure}[ht]
	$$
	\includegraphics[scale=0.6]{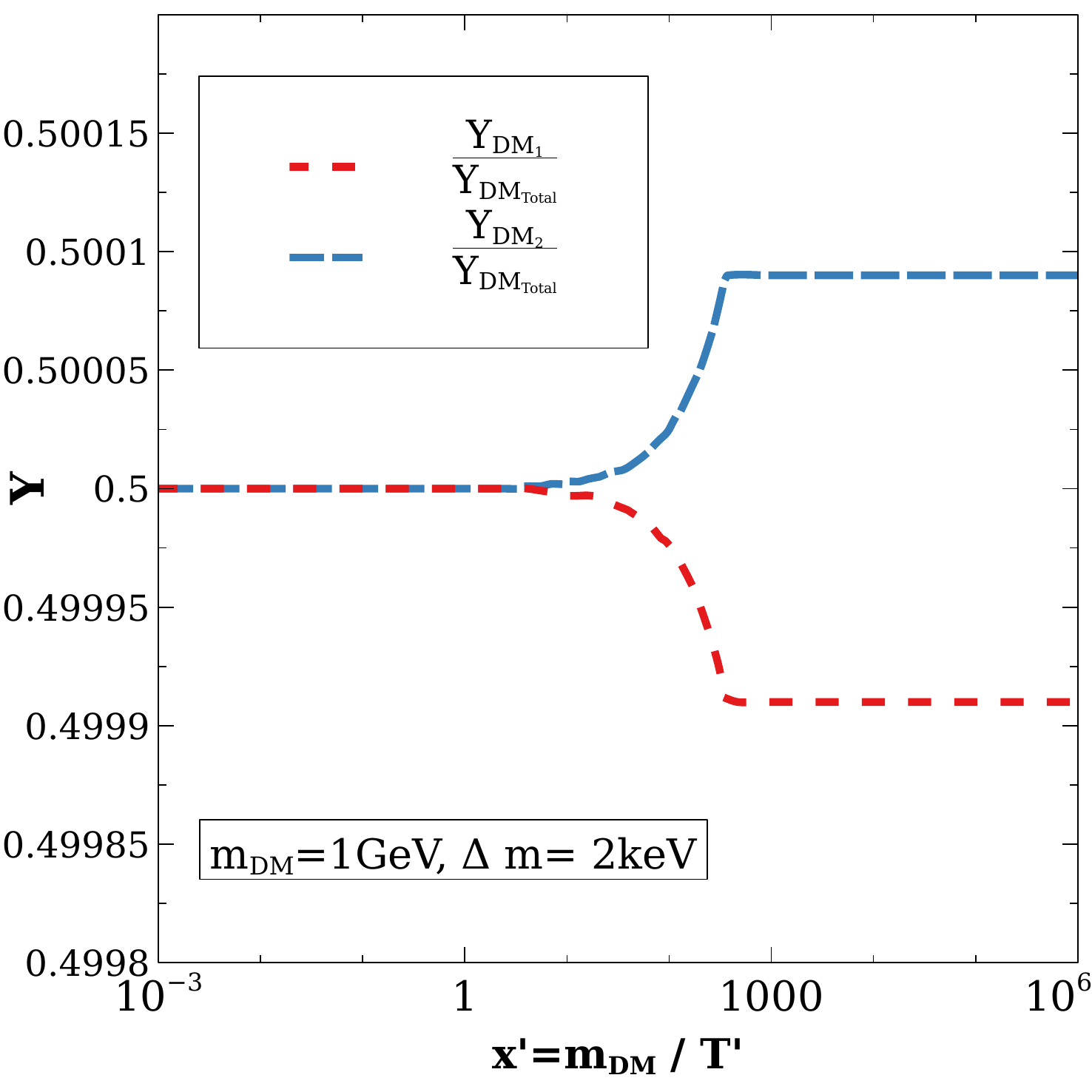} 
	$$ 
	\caption{\footnotesize{Fractional contributions $Y_{\rm DM_1}/Y_{\rm DM_{\rm Total}}$ and  $Y_{\rm DM_2}/Y_{\rm DM_{\rm Total}}$ to DM relic density for $\Delta m = 2\times 10^{-6}$ GeV}.}
	\label{relic_components}
\end{figure}

Note that the lines showing the evolution of DM number density in figure~\ref{relic} considers both the DM components $\psi_{1,2}$. Since their mass splitting is very small $\Delta m \sim \mathcal{O}(\rm keV)$ they behave very similarly as far as calculation of relic abundance goes. However, once the net relic is generated, there can be interconversion between two DM components dominantly through $Z'$-mediated t-channel process $\psi_2 \psi_2 \rightarrow \psi_1 \psi_1$. We take this into account and show that the effect of such interconversion with such small mass splitting ($\Delta m = 2\times 10^{-6}$ GeV) is negligible. This can be seen from figure~\ref{relic_components}, where the fractional contributions $Y_{\rm DM_1}/Y_{\rm DM_{\rm Total}}$ and  $Y_{\rm DM_2}/Y_{\rm DM_{\rm Total}}$ for mass splitting $\Delta m = 2\times 10^{-6}\; {\rm GeV}$ are shown. We have also taken into account the Sommerfeld effect induced by the multiple $Z'$ boson exchange in the inter-conversion process \cite{Slatyer:2009vg}. Clearly, such interconversions lead to negligible effects on individual DM relic abundance and hence we consider them to be equally dominant in rest of our analysis.

\section{The XENON1T Excess}\label{xenon1t}
\begin{figure}[hb]
	\centering
	\includegraphics[scale=0.5]{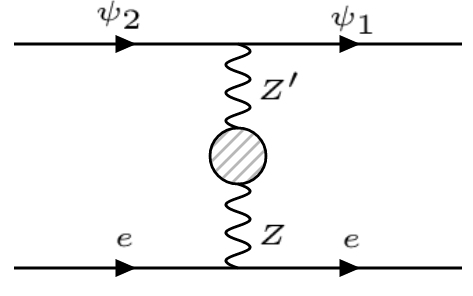}
	\caption{\footnotesize{DM-electron scattering at XENON1T}}
\end{figure}

The direct detection prospects of such self-interacting DM can be addressed through the recently reported excess in the electron recoil events at XENON1T experiment. We assume $\psi_2$ is heavier than $\psi_1$ with a small mass splitting $\Delta m = M_2-M_1$ 
between the two components. Because of this inelastic nature of these DM candidates and since the mass splitting $\Delta m$ is kept fixed at keV scale, it can successfully explain the recently reported XENON1T anomaly \cite{Aprile:2020tmw}. For a fixed incoming velocity $v$ of heavier DM $\psi_2$, the differential 
scattering cross section for the down scattering process $\psi_2 e \rightarrow 
	\psi_1 e$ (with electrons inside the Xenon atom) can be written as
	\begin{equation}
		\frac{d \langle \sigma v \rangle }{d E_r} = \frac{\sigma_e}{2 m_e v } \int_{q-}^{q+} a^2_0 q dq|F(q)|^2 K(E_r,q)\,,
		\label{Event_central}
	\end{equation}	
	where $m_e$ is the electron mass, $\sigma_e$ is the corresponding free electron cross section at fixed momentum transfer 
	$q=1/a_0$ with $a_0 = \frac{1}{\alpha m_e}$ being the Bohr radius, $\alpha = \frac{e^2}{4 \pi}=\frac{1}{137}$ being 
	the fine structure constant, $E_r$ is the recoil energy of electron and $K(E_r, q)$ is the atomic excitation factor. For our calculations, the 
atomic excitation factor is adopted from \cite{Roberts:2019chv}. We assume the DM form factor to be unity. 

However, to include velocity dispersion in Eq. \eqref{Event_central}, we use the following distribution function (obtained 
after angular integration of a Maxwellian velocity distribution boosted in earth's rest frame)
\begin{equation}\label{max_distribution}
f(v)= A v {\rm Exp}[-3(v-v_m)^2/2\sigma^2_v]\,,
\end{equation}
where $A$ is the normalisation constant such that $\int f(v) dv=1$. The details of velocity distribution is given in Appendix~\ref{v_distribution}. In Eq. \eqref{max_distribution}, $v_m$ is the most-probable velocity of DM which is induced by the relative velocity of the Sun w.r.t galactic halo.
Here $\sigma_v$ is the DM velocity dispersion which is given by $\sigma^2_v= \frac{3}{2}v^2_m$. As a result Eq. \eqref{Event_central}, after incorporating velocity dispersion of DM, can be rewritten as \cite{Roberts:2016xfw,Roberts:2019chv,Bramante:2020zos,He:2020wjs}
\begin{equation}
		\frac{d \langle \sigma v \rangle }{d E_r} = \frac{\sigma_e}{2 m_e }\int_{0}^{v_{esc}} dv \frac{f(v)}{v} \int_{q-}^{q+} a^2_0 q dq|F(q)|^2 K(E_r,q)\,.
		\label{Event}
	\end{equation}	
Where $v_{esc}$ is the DM escape velocity in the Milky Way which is of the order $v_{esc} \sim$ $533^{+54}_{-41}$ km/s\cite{Piffl:2013mla}. In inelastic DM scenarios, the minimum DM velocity ($v_{min}$) required by the DM to upscatter to the NLSP and register a recoil inside the detector is decided by the kinematics of scattering. However, it is worth mentioning that in the case of an inelastic down scattering of DM with electron, which we consider here, there is no kinematic limit on the minimum velocity of DM as the incoming particle with almost vanishing velocity can still down scatter to the lighter component with the mass
splitting between the DM components being transferred to the electron recoil energy, without violating anything kinematically.

The free electron scattering cross-section for the process $\psi_2 e \rightarrow \psi_1 e$ is given by
\begin{equation}
	\sigma_e = \frac{16 \pi \alpha_Z \alpha^{'} \epsilon^2 m^2_e  }{M^4_{Z'}}
	\label{DM-electron-scattering}
\end{equation}
where $\alpha_Z=\frac{g^2}{4 \pi}$, $\alpha^{'}=\frac{g'^2}{4 \pi}$ and $\epsilon$ is the kinetic mixing parameter between $Z$ and $Z'$ gauge bosons. For chosen values of DM and mediator masses in our work, this kinetic mixing is required to be $\epsilon \sim 10^{-8}$. It should be noted that, for GeV scale DM, $\sigma_e$ 
is independent of DM mass as the reduced mass of DM-electron is almost equal to electron mass. 
The limits of integration for the inelastic scattering in Eq.~\eqref{Event} are determined depending on the relative values of recoil energy ($E_r$) and the mass splitting between the two DM components. 

For $E_r \geq \Delta m$
\begin{equation}
	q_\pm=M_{2} v \pm \sqrt{M^2_{2} v^2 -2M_{2}(E_r-\Delta m)}\,.
\end{equation}
And for $E_r \leq \Delta m$
\begin{equation}
	q_\pm=\sqrt{M^2_{2} v^2 -2M_{2}(E_r-\Delta m)} \pm M_{2} v \,.	
\end{equation}

\begin{figure}
	\centering
	\includegraphics[scale=0.4]{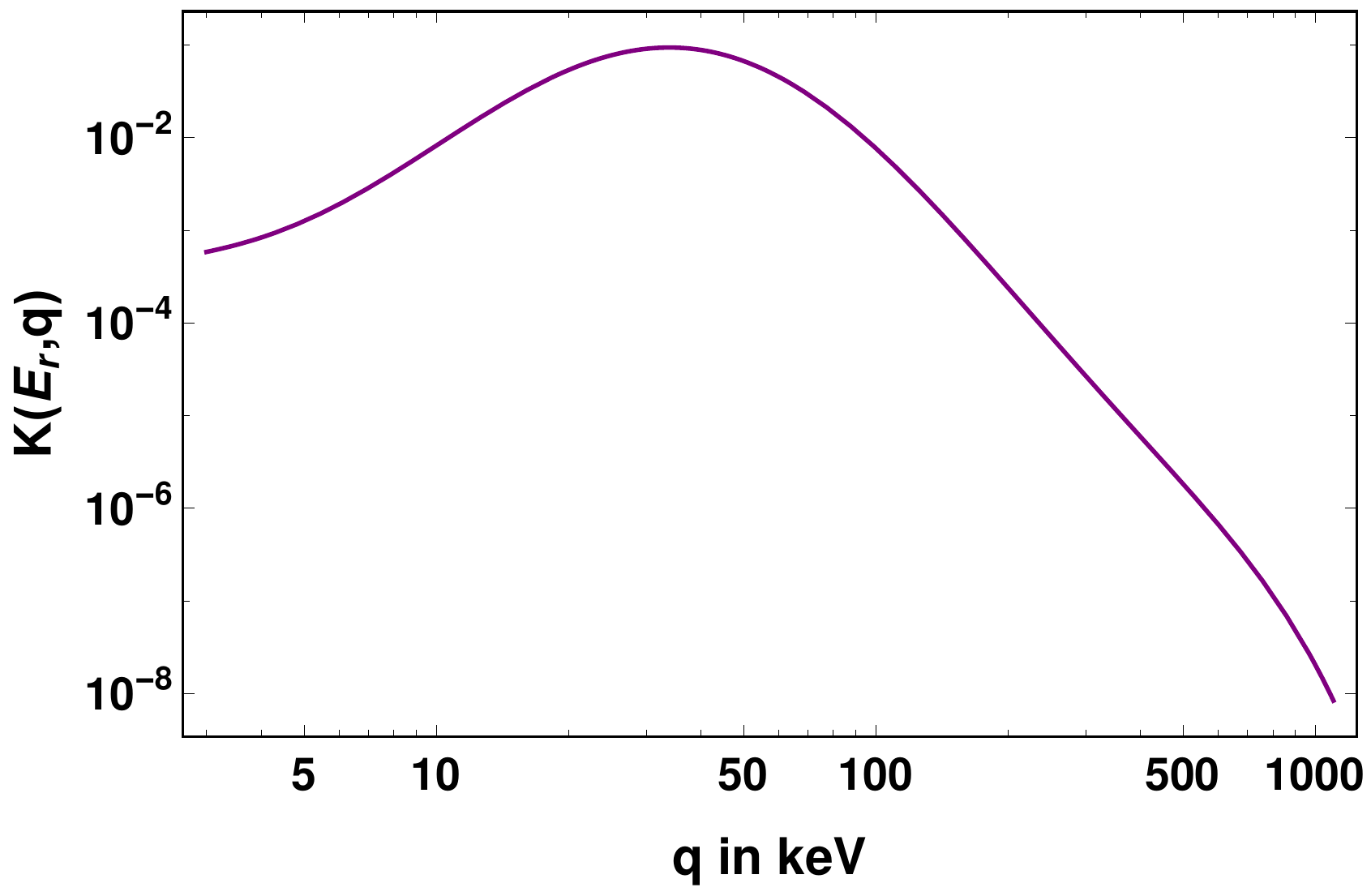}	
	\hfil
	\includegraphics[scale=0.45]{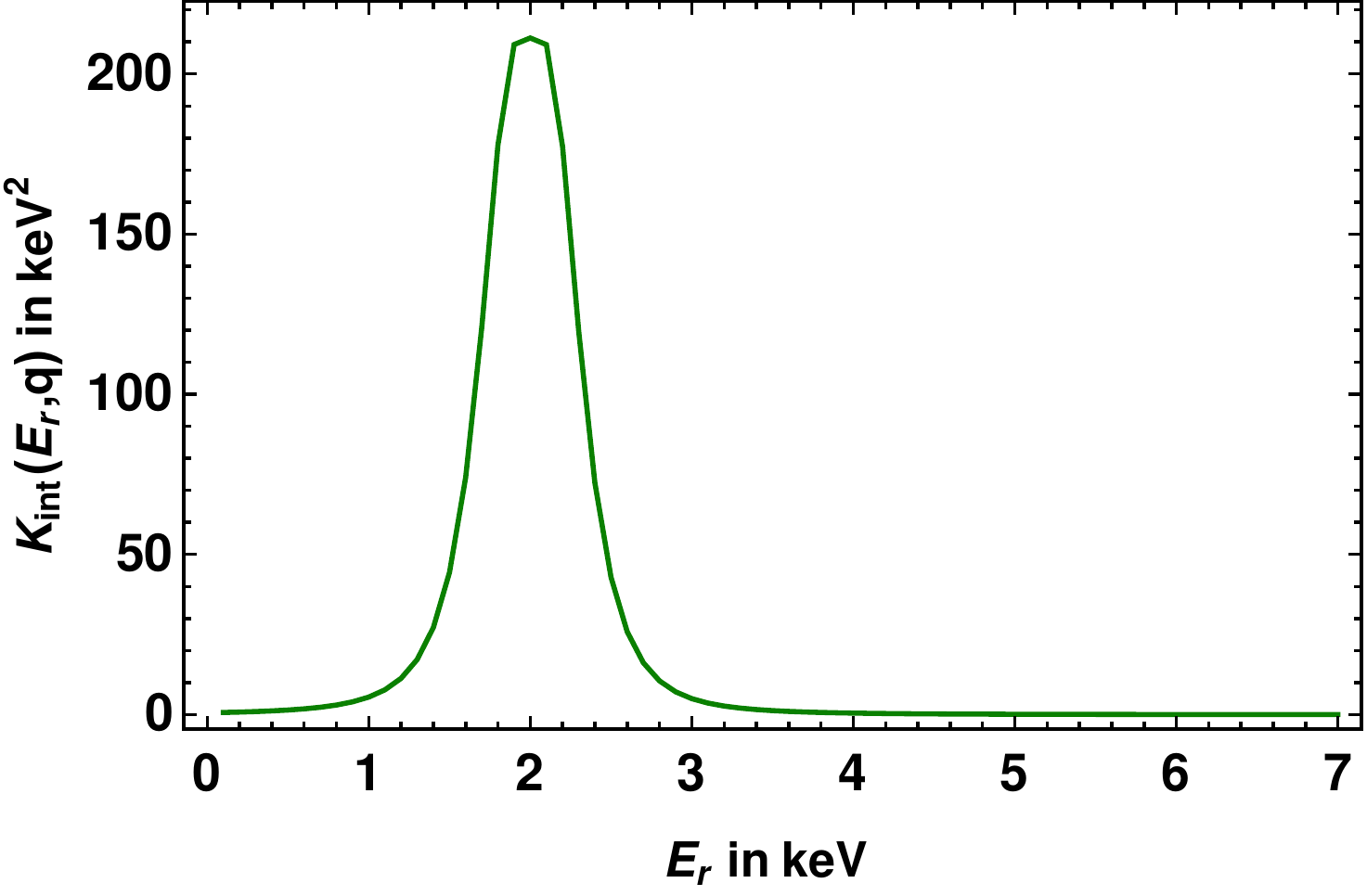}
	\caption{\footnotesize{Left panel: Atomic excitation factor is shown as a function of momentum transferred. Right panel: The atomic excitation factor after being integrated over the transferred momentum, is shown as a function of the transferred recoil energy $E_r$.}}
	\label{aef}
\end{figure}

The dependency of atomic excitation factor on the momentum transferred $q$ is shown in figure~\ref{aef}. Here the dominant contribution comes from the bound states with principal quantum number $n=3$ as their binding energy is around a few 
keVs. In the right panel of figure~\ref{aef}, we have shown the plot for the integration of momentum transferred times the 
atomic excitation factor $\big({\it i.e.}K_{\rm int}(E_r,q)=\int_{q-}^{q+} q dq K(E_r,q)\big)$ as a function of the recoil energy 
$E_r$ for $M_1=0.3$GeV and $\Delta m = 2$keV. The figure shows a peak around $E_r \simeq \Delta m$ since the $q_{-}$ 
approaches to zero and the momentum transfer maximising this factor is available. It is worth mentioning that such kind 
of enhancement is a characteristic feature of inelastic scattering.

The differential event rate for the inelastic DM scattering with electrons in Xenon atom, {\it i.e} $\psi_2 e 
\rightarrow \psi_1 e$, can be given as: 
\begin{equation}
	\frac{dR}{dE_r}=n_T n_{\rm DM} \frac{d \langle \sigma v \rangle }{d E_r}
	\label{event_rate}
\end{equation}
\begin{figure}
	\centering
	\includegraphics[scale=0.5]{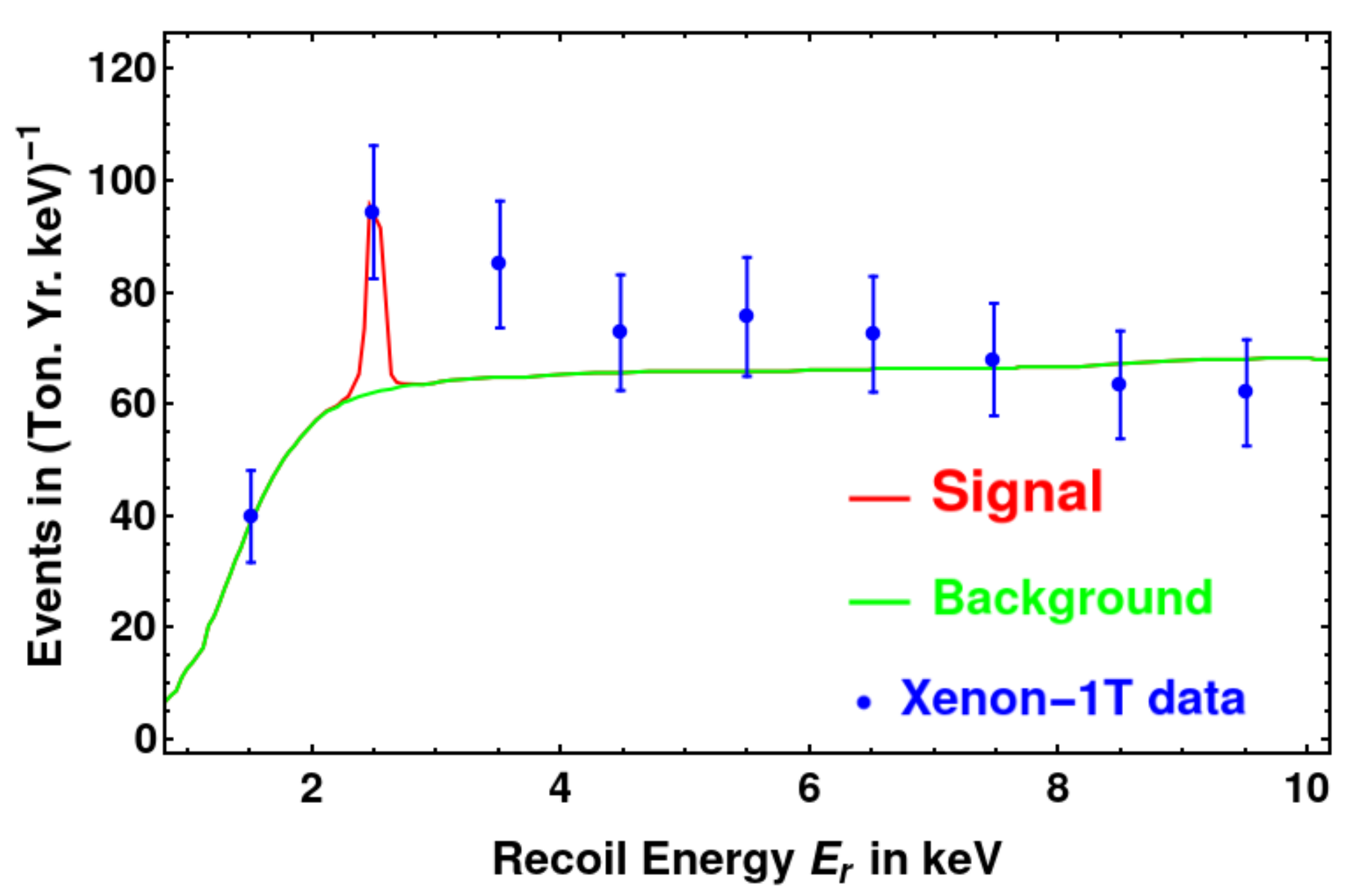}
	\includegraphics[scale=0.5]{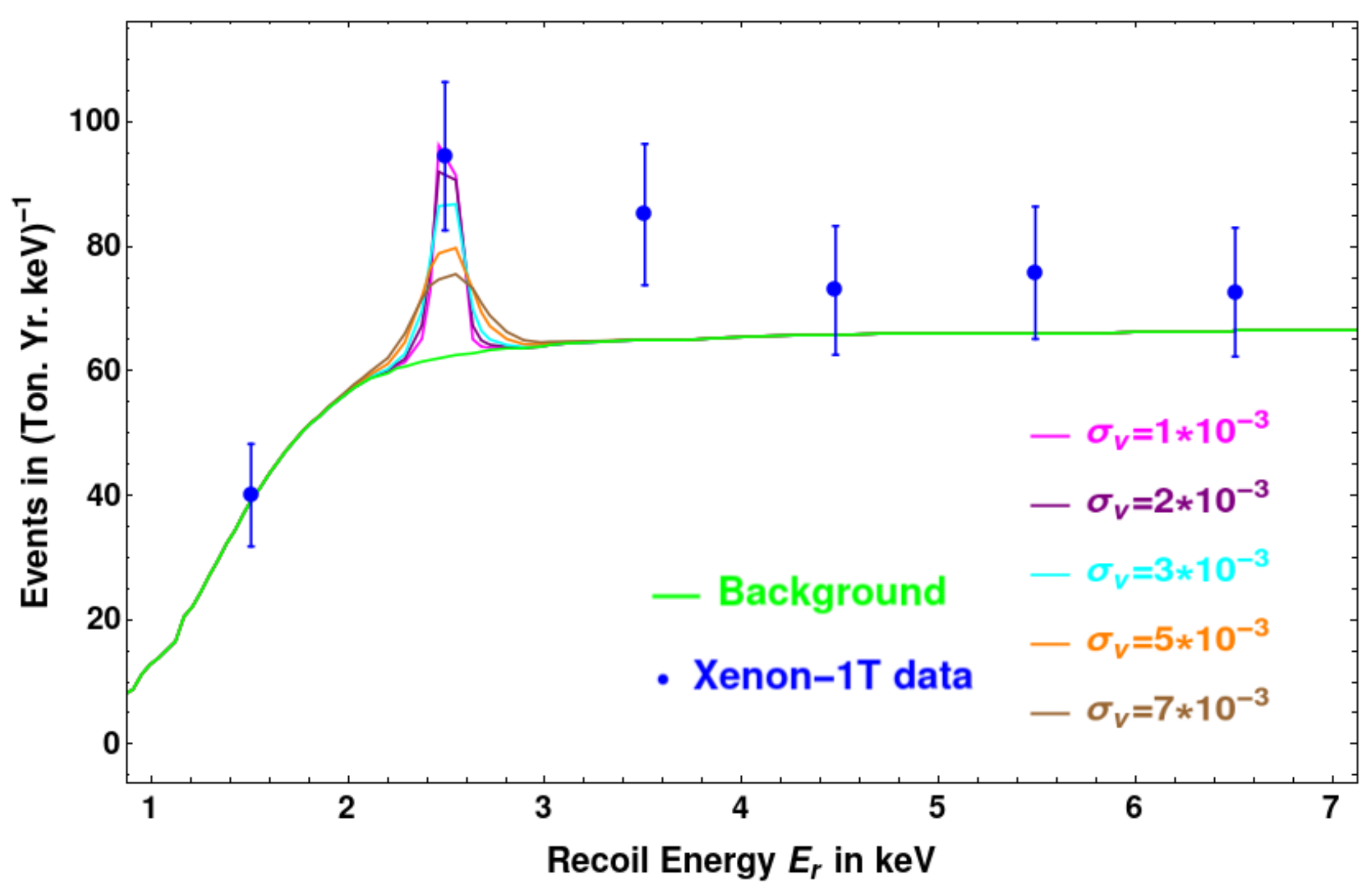}	
	\caption{\footnotesize{Fit to XENON1T electron recoil excess with the self interacting inelastic DM in our model.}}
			
	\label{xenonfit}
\end{figure}
where $n_T=4\times10^{27}$ $ {\rm Ton}^{-1}$ is the number density of Xenon atoms and $n_{\rm DM}$ is the number density of the dark matter particle.

The detected recoil energy spectrum can be obtained by convolving Eq.~\eqref{event_rate} with the energy resolution of the XENON1T 
detector. Incorporating the detector efficiency $\gamma(E)$, the energy resolution of the detector is given by a Gaussian distribution with an 
energy dependent width, 
\begin{equation}
	\zeta(E,E_r)=\frac{1}{\sqrt{2 \pi \sigma^2_{\rm det}}}{\rm Exp}\Big[-\frac{(E-E_r)^2}{2 \sigma^2_{\rm det}}\Big] \times \gamma(E)
\end{equation}
where $\gamma(E)$ is reported in figure~2 of \cite{Aprile:2020tmw} and the width $\sigma_{\rm det}$ is given by 
\begin{equation}
\sigma_{\rm det}(E)= a \sqrt{E} + b E
\end{equation} 
with $a=0.3171$ and $b=0.0037$.
Thus the final detected recoil energy spectrum is given by
\begin{equation}
\frac{dR_{\rm det}}{dE_r}=\frac{n_T n_{\rm DM} \sigma_e a^2_0}{2 m_e }  \int dE ~~\zeta(E,E_r) \Bigg[\int_{0}^{v_{esc}} dv \frac{f(v)}{v} \int_{q-}^{q+} dq~~ q K(E_r,q)\Bigg] 	
\end{equation}

To obtain the fit shown in the upper panel of figure~\ref{xenonfit}, the mass splitting is taken to be $\Delta m = 2$ keV while heavier DM mass is taken to be 1 GeV. The other relevant parameters used in this fit are $\sigma_v=\sqrt{3/2}v_m$ with $v_m = 1 \times 10^{-3}$, $g' =0.1$, $M_{Z'} =10$MeV, 
$ \epsilon = 4 \times 10^{-8}$ which corresponds to cross section $\sigma_e = 1.9 \times 10^{-17} \; {\rm GeV}^{-2}$. 

On the other hand, in the bottom panel of figure~\ref{xenonfit}, we have shown the fit considering different velocity dispersion for the DM particle 
as we have no observational constraints on $f(v)$ apart from numerical simulations. Clearly as we increase the velocity dispersion the peak in the spectrum giving an appreciable fit gets flatten out and no longer explain the XENON1T signal within $E_r=2-3$keV for larger $\sigma_v$. 


\section{Summary and Conclusion}
\label{conclude}
\begin{figure}{b}
	\centering
	\includegraphics[scale=0.85]{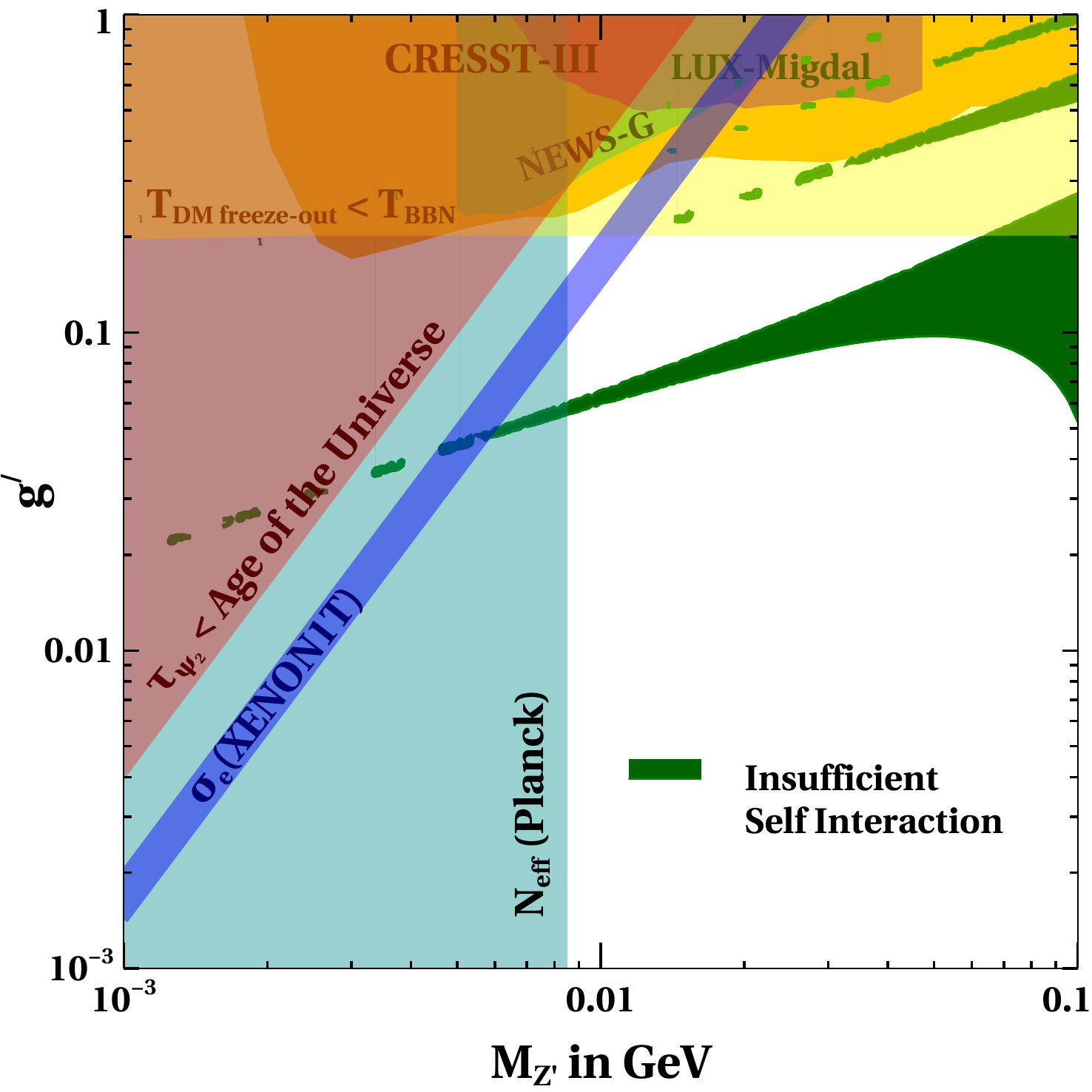}	
	\caption{\footnotesize{Summary plot for inelastic self-interacting DM showing the final parameter space from relevant constraints. The white region represents the allowed parameter space available after imposing all the constraints. The blue patch represents the parameter space allowed by XENON1T for 1 GeV inelastic DM with mass splitting $\Delta m=2$ keV and kinetic mixing parameter $ \epsilon = 4\times 10^{-8}$.  }}
	\label{DIDM_summary}
\end{figure}

We summarise our key findings in figure~\ref{DIDM_summary}. We show all the relevant constraints as well as favoured parameter space in the $g'-M_{Z'}$ plane. In figure~\ref{DIDM_summary}, all the coloured regions (except the blue one which favoured from XENON1T excess) represent disfavoured regions from different bounds. The green patch represents the region where the DM self-scattering cross-section is not large enough to solve the astrophysical problems discussed in section \ref{sec3}. To be more quantitative, the green shaded regions correspond to DM self-scattering cross-section $\sigma/m < 0.1 \; {\rm cm}^2/{\rm g}$. The triangular region on upper left corner of figure~\ref{DIDM_summary} is disfavoured from lower bound on lifetime of heavier DM. Since the mass splitting between $\psi_1$ and $\psi_2$ is kept at keV scale $\Delta m= \mathcal{O}(keV)$, there can be decay modes
like $\psi_2 \rightarrow \psi_1 \nu \overline{\nu}$  mediated by $Z-Z'$ mixing. If both the DM components are to be there in the present 
universe, this lifetime has to be more than the age of the universe, that is $\tau_{\psi_2} > \tau_{\rm Univ.}$. The decay width of this 
process is $\Gamma(\psi_2 \rightarrow \psi_1 \nu \overline{\nu})= \frac{g^2 g'^2 \epsilon^2 (\Delta m)^5}{160 \pi^3 M^4_{Z'}}$. Thus, imposing 
the lifetime constraint on heavier DM, we get the triangular shaded region.
We also show the parameter space excluded by 
	the recent results from CRESST-III~\cite{Abdelhameed:2019hmk}, LUX-Migdal~\cite{Akerib:2018hck} and NEWS-G~\cite{Arnaud:2017bjh} on low mass DMs . This corresponds to the shaded region of orange,brown and light green colour at topmost part of figure~\ref{DIDM_summary}. The bound from EDELWEISS-III~\cite{Armengaud:2019kfj} is much weaker than the above mentioned experiments. Assuming $M_{Z'}=0.01 M_{DM}$, these are the only experiments that are sensitive to the parameter space we are interested in. We have checked that the constraints from other low-threshold experiments like DAMIC, PICO, PANDAX-II, CDMSlite etc. do not apply to our parameter space.  The solid band of blue colour corresponds to free electron cross section $\sigma_e=(1 - 5)\times10^{-17}$ 
GeV$^{-2}$ which is required to obtain the fit for the XENON1T excess for a DM of mass around 1 GeV with a typical DM velocity of order $\mathcal{O}(10^{-3})$. The shaded region of yellow colour at top corresponds to the region where DM annihilation into $Z'$ pairs does not freeze out before the epoch of big bang nucleosynthesis (BBN). This will require scalar singlet decay at post-BBN epochs. Additionally the $Z'$ bosons which keep getting produced from DM annihilations will decay into light SM fermions injecting new relativistic degrees of freedom. Since all these may potentially ruin the successful predictions of the BBN, we disfavour this region of parameter space. Since our chosen value of kinetic mixing is very small, the flavour bounds on such light $Z'$ bosons from dark photon searches at BABAR \cite{Lees:2014xha} are automatically satisfied. Additionally, CMB bounds from Planck measurements on DM annihilations into charged fermions \cite{Aghanim:2018eyx} are trivially satisfied as all such processes remain suppressed by kinetic mixing. Another constraint on the parameter space arise due to late decay of $Z'$ into SM leptons. For example, if $Z'$ decays after neutrino decoupling temperature $T^{\nu}_{\rm dec} \sim \mathcal{O}(\rm MeV)$, it will increase the effective relativistic degrees of freedom which is tightly constrained by Planck 2018 data as ${\rm N_{eff}= 2.99^{+0.34}_{-0.33}}$ \cite{Aghanim:2018eyx}. As pointed out by the authors of \cite{Ibe:2019gpv}, such constraints can be satisfied if $M_{Z'} \gtrsim 8.5 \; {\rm MeV}$ for the chosen value of kinetic mixing parameter in our work. We show this as the light green shaded region towards left in figure~\ref{DIDM_summary}. Note that we have not imposed any constraints from DM relic point of view as that can be satisfied independently by appropriate tuning of scalar singlet parameters discussed before. 

To conclude, we have studied the possibility of self-interacting DM as a possible explanation of the recently reported XENON1T excess. While XENON1T excess can arise due to inelastic nature of DM so that the heavier DM can undergo a down scattering with electrons, the corresponding mediator of such scattering, if sufficiently light compared to DM can also give rise to the required self-interaction cross section $\sigma/m$ required to solve the small scale structure problems associated with cold dark matter. We consider a hidden $U(1)_X$ gauge symmetry under which the inelastic DM is charged and this dark sector interacts with the SM purely via kinetic mixing of $U(1)_X$ with $U(1)_Y$ of the standard model. The requirement of large self-interaction or $U(1)_X$ gauge coupling forces us to consider tiny kinetic mixing required to generate the XENON1T excess while satisfying all other experimental bounds. This tiny kinetic mixing also prevents DM from reaching chemical equilibrium with the SM requiring its non-thermal or freeze-in production from the SM bath. However, due to large coupling of DM with $U(1)_X$ gauge boson $Z'$, they can annihilate strongly into much lighter $Z'$ bosons depleting the number density generated from freeze-in. To fill the gap, we introduce another long-lived scalar singlet which freezes out from the thermal bath and decays very late into DM generating the required relic. As seen from the summary plot in figure \ref{DIDM_summary}, after applying all relevant bounds, there exists only a tiny parameter space (the blue shaded region not overlapped with other regions) that can give rise to the required XENON1T excess, DM self-interactions for 1 GeV inelastic DM with mass splitting of 2 keV while being consistent with all other bounds. Future data from XENON1T experiment as well as other searches should be able to further constrain or confirm this predictive scenario. 

Now we turn to comment on the implications of thermally generated self interacting dark matter $\psi_1$ and $\psi_2$, which we 
assume to constitute about $1\%$ of the total relic (see figure \ref{Thermal_relics}). Since the relic density is smaller by two orders of magnitude than the observed one, 
the corresponding DM-electron cross-section $\sigma_e (\psi_2 e \to \psi_1 e)$ has to be increased by two orders in order to explain the observed 
XENON1T excess. This can be achieved by increasing $\epsilon$ by one order of magnitude, since $\sigma_e \propto \epsilon^2$. However, increasing 
$\epsilon$ by one order of magnitude will not satisfy the lifetime bound on $\psi_2$ as $\tau_{\psi_2} \propto 1/\epsilon^2$. Note that such 
sub-dominant SIDM will not solve the small scale structure problem as well, even if DM deficit is filled by some other component which is neither connected to 
SIDM nor to the observed XENON1T excess.

\acknowledgements
DB acknowledges the support from Early Career Research Award from DST-SERB, Government of India (reference number: ECR/2017/001873). MD acknowledges Department of Science and Technology (DST), Govt. of India for providing the financial assistance for the research under the grant 
DST/INSPIRE/03/ 2017/000032. MD would also like to acknowlege Tracy R. Slatyer and Katelin Schutz for useful discussion regarding self-interating dark matter. 

\appendix

\section{Relevant cross section and decay widths}

\subsection{Self-interaction cross sections at low energy}
\label{appen1}
The scattering cross sections can be derived as \cite{Schutz:2014nka}

\begin{equation}
	\sigma_{\psi_1 \psi_1 \rightarrow \psi_1 \psi_1}=\frac{\pi}{\epsilon^2_v}\Bigg|1+\Bigg(\frac{V_0}{4\mu^2}\Big)^{-\frac{2i\epsilon_{v}}{\mu}}\Big(\frac{\Gamma_v}{\Gamma^*_v}\Big)\Bigg[\frac{\cosh(\frac{\pi(\epsilon_\Delta+\epsilon_v)}{2\mu})\sinh(\frac{\pi(\epsilon_v-\epsilon_\Delta)}{2\mu}+i\varphi)}{\cosh(\frac{\pi(\epsilon_\Delta-\epsilon_v))}{2\mu}\sinh(\frac{\pi(\epsilon_v+\epsilon_\Delta)}{2\mu}-i\varphi)}\Bigg]\Bigg|^2
\end{equation}

\begin{equation}
	\sigma_{\psi_2 \psi_2 \rightarrow \psi_2 \psi_2}=\frac{\pi}{\epsilon^2_\Delta}\Bigg|1+\Bigg(\frac{V_0}{4\mu^2}\Big)^{-\frac{2i\epsilon_{\Delta}}{\mu}}\Big(\frac{\Gamma_\Delta}{\Gamma^*_\Delta}\Big)\Bigg[\frac{\cosh(\frac{\pi(\epsilon_\Delta+\epsilon_v)}{2\mu})\sinh(\frac{\pi(\epsilon_v-\epsilon_\Delta)}{2\mu}+i\varphi)}{\cosh(\frac{\pi(\epsilon_\Delta-\epsilon_v))}{2\mu}\sinh(\frac{\pi(\epsilon_v+\epsilon_\Delta)}{2\mu}-i\varphi)}\Bigg]\Bigg|^2
\end{equation}

\begin{equation}
	\sigma_{\psi_1 \psi_1 \rightarrow \psi_2 \psi_2}=\frac{2\pi \cos^2\varphi \sinh\Big(\frac{\pi \epsilon_v)}{\mu}\Big)\sinh\Big(\frac{\pi \epsilon_{\Delta}}{\mu}\Big)}{\epsilon^2_v \cosh^2\Big(\frac{\pi(\epsilon_{\Delta} - \epsilon_v)}{2\mu}\Big)\Big(\cosh\Big(\frac{\pi(\epsilon_v+\epsilon_\Delta)}{\mu}\Big)-\cosh(2\varphi)\Big)}
\end{equation}

\begin{equation}
	\sigma_{\psi_2 \psi_2 \rightarrow \psi_1 \psi_1}=\frac{2\pi \cos^2\varphi \sinh\Big(\frac{\pi \epsilon_v)}{\mu}\Big)\sinh\Big(\frac{\pi \epsilon_\Delta}{\mu}\Big)}{\epsilon^2_\Delta \cosh^2\Big(\frac{\pi(\epsilon_{\Delta} - \epsilon_v)}{2\mu}\Big)\Big(\cosh\Big(\frac{\pi(\epsilon_v+\epsilon_\Delta)}{\mu}\Big)-\cosh(2\varphi)\Big)}
\end{equation}

where we have defined, $\epsilon_\Delta=\sqrt{\epsilon^2_v-\epsilon^2_\delta}$, $\mu$ and $V_0$ are defining parameters for the exponential potential $V_0e^{-\mu r}$, given by,
\begin{equation}
	\mu=\epsilon_Z\Bigg(\frac{1}{2}+\frac{1}{2}\sqrt{1+ \frac{4}{\epsilon_Z r_M}}\Bigg),  ~~~~~V_0=\frac{e^{\epsilon_Z r_M\Big(-\frac{1}{2}+\frac{1}{2}\sqrt{1+\frac{4}{\epsilon_Z r_M}}\Big)}}{r_M}.
\end{equation}
Here $r_M$ is chosen from the relation $e^{-\epsilon_{\phi} r_M}/r_M = {\rm max}( \epsilon^2_{\delta}/2, \epsilon^2_{\phi})$. The terms $\Gamma_v, \Gamma_{\Delta}$ are given by 
\begin{equation}
	\Gamma_v = \Gamma \left ( 1+ i \frac{\epsilon_v}{\mu} \right) \Gamma \left( i \frac{\epsilon_v-\epsilon_{\Delta}}{2\mu}+\frac{1}{2} \right) \Gamma \left( i \frac{\epsilon_v+\epsilon_{\Delta}}{2\mu}+\frac{1}{2} \right)
\end{equation}
\begin{equation}
	\Gamma_{\Delta} = \Gamma \left ( 1+ i \frac{\epsilon_{\Delta}}{\mu} \right) \Gamma \left( i \frac{\epsilon_{\Delta}-\epsilon_{v}}{2\mu}+\frac{1}{2} \right) \Gamma \left( i \frac{\epsilon_v+\epsilon_{\Delta}}{2\mu}+\frac{1}{2} \right)
\end{equation}
with $\Gamma$ denoting the gamma function.

\subsection{Interactions for DM relic calculations}
\label{appen2}

\begin{eqnarray}
	\sigma({\rm DM\; DM} \rightarrow e^{+} e^{-}) &=& \frac{g^2 g'^2 \epsilon^2 (2s+(M_{\psi_1}+M_{\psi_2})^2)\sqrt{M^4_{\psi_1} + (s-M^2_{\psi_2})^2 - 2M^2_{\psi_1}(s+M^2_{\psi_2})}}{192 \pi \cos^2 \theta_{W}(s-M^2_{Z'})^2 (s-(M_{\psi_1}+M_{\psi_2})^2)}  \nonumber \\
	\sigma({\rm DM \;DM} \rightarrow Z' Z') &\simeq& \frac{g'^4}{192 \pi M^4_{Z'} s (s-4M^2_{\psi})} \times \Bigg[\frac{24M^4_{Z'}s(4m^4_{\psi}+2M^4_{Z'}+sM^2_{\psi})A}{M^4_{Z'}+M^2_{\psi}(s-4M^2_{Z'})}\nonumber\\ & -&\frac{24 M^4_{Z'}(8M^2_{\psi}-4M^2_{Z'}-s^2-(s-2M^2_{Z'})4M^2_{\psi})}{s-2M^2_{Z'}} {\rm Log}\Big[\frac{2M^2_{Z'}+s(A-1)}{2M^2_{Z'}-s(A+1)}\Big]\Bigg]\nonumber
\end{eqnarray}
where $A=\sqrt{\frac{(s-4M^2_{Z'})(s-4M^2_{\psi_1})}{s^2}}$
\begin{eqnarray}
	\sigma( e^{+} e^{-} \rightarrow {\rm DM \;DM})&=& \frac{g^2 g'^2 \epsilon^2(s+2M^2_{\psi})(s-M^2_e-4(s+2M^2_e)\sin^2\theta_{W})}{96 \pi \cos^2\theta_{W}(s-4M^2_e)(s-M^2_{Z'})^2}\sqrt{\frac{(s-4M^2_e)(s-4M^2_\psi)}{s^2}} \nonumber \\
	\sigma( {\rm DM}  e^{-} \rightarrow {\rm DM}  e^{-})&=& \frac{g^2 g'^2 \epsilon^2  A}{128\pi \cos^2\theta_{W} m^4_{Z'}(s-M^2_e - M^2_{\psi_2})^2} \frac{B}{C}- D  {\rm Log} \Big[\frac{E+s(2 M^2_{Z'} - M^2_{\psi_2} + s+A)}{E+s(2 M^2_{Z'} - M^2_{\psi_2} + s-A)}\Big] \nonumber
\end{eqnarray}
where
\begin{eqnarray}
	A&=&s\Big({\frac{(M^4_e + (M^2_{\psi_1} - s)^2 - 2 M^2_e (M^2_{\psi_1} + s)) (M^4_e + (M^2_{\psi_2} - s)^2 -	2 Me^2 (M^2_{\psi_2} + s))}{s^4}}\Big)^{\frac{1}{2}} \nonumber \\
	B&=&s (2 M^4_{Z'} s + 2(M^2_{\psi_1} - s) (M^2_{\psi_2} - s) s + M^4_e (M^2_{Z'} + 2 s) + 
	M^2_{Z'} (M^2_{\psi_1} M^2_{\psi_2}  
	\nonumber\\&-&2 (M^2_{\psi_1} - M_{\psi_1} M_{\psi_2} + M^2_{\psi_2}) s + 3 s^2))\nonumber\\&+&M^2_e ((M^2_{\psi_1} - M^2_{\psi_2})^2 + 4 M_{\psi_1}M_{\psi_2} s - 4 s^2 - 
	M^2_{Z'} (M^2_{\psi_1} + M^2_{\psi_2} + 2 s))\nonumber\\
	C&=&M^4_e M^2_{Z'} + M^2_{Z'} ((M^2_{\psi_1} - s) (M^2_{\psi_2} - s) + M^2_{Z'} s) + 
	M^2_e ((M^2_{\psi_1} - M^2_{\psi_2})^2 - 
	M^2_{Z'} (M^2_{\psi_1} + M^2_{\psi_2} + 2 s))\nonumber\\
	D&=&2 M^2_{Z'} - (M_{\psi_2} - M_{\psi_1})^2 + 2 s\nonumber\\
	E&=&M^4_e + M^2_{\psi_1} (M^2_{\psi_2} - s) - 
	M^2_e (M^2_{\psi_1} + M^2_{\psi_2} + 2 s)\nonumber
\end{eqnarray}

\begin{equation}
	\sigma (\eta^\dagger \eta \to H^\dagger H)= \frac{\lambda}{16\pi s}\sqrt{\frac{s-4 M^2_\eta}{s-4 M^2_H}} 
\end{equation}

The decay width of the scalar singlet $\eta$ is given by:
\begin{equation}
	\Gamma({\eta\rightarrow DM DM}) = \frac{\lambda^2}{8\pi}m_{\eta}\Big(1-4\frac{m^2_{DM}}{m^2_\eta} \Big)^{3/2}
\end{equation}

The decay width of $Z'$ is given by:
\begin{equation}
	\Gamma(Z'\rightarrow f \bar{f}) = \frac{\epsilon^2 g^2 M_{Z'}}{48 \pi \cos^2 \theta_{W}} (C^2_{V_f}+C^2_{A_f}) 
\end{equation}

Thermal average cross-section is given by \cite{Gondolo:1990dk}
\begin{equation}
	\langle\sigma v \rangle_{CM} = \frac{x}{2\big[K^2_1(x)+K^2_2(x)\big]}\times \int^{\infty}_{2}  dz \sigma (z^2 m^2_\psi) (z^2 - 4)z^2 K_1(zx)
\end{equation}

\subsection{DM Velocity distribution function}
\label{v_distribution}
The distribution function used in Eq. \ref{max_distribution} can be obtained as follows. Let $~\overrightarrow{u}$ and $\overrightarrow{v}$ are 
the velocities of dark matter in the rest frames of galaxy and earth respectively. If $\overrightarrow{v_E}$ is the velocity of earth with 
respect to the galactic rest frame then we have $\overrightarrow{u} = \overrightarrow{v}+\overrightarrow{v_E}$. Assuming that the velocity 
distribution of dark matter with respect to the galactic rest frame is Maxwellian, we can write 
\begin{eqnarray}
	f(\overrightarrow{u})d^3u =N~ e^{\frac{-3|\overrightarrow{u}|^2}{2 \sigma^2}} d^3 u = N ~e^{\frac{-3(\overrightarrow{v_E}+\overrightarrow{v})^2}{2 \sigma^2_v}} d^3 v
\end{eqnarray}
where N is the normalisation constant and $\sigma_v$ is the velocity dispersion. Assuming spherical symmetry and considering $z$-axis in the direction of 
$\overrightarrow{v_E}$ which subtends an angle $\theta$ with $\overrightarrow{v}$, we can write :
\begin{align}
	N~e^{\frac{-3(\overrightarrow{v_E}+\overrightarrow{v})^2}{2 \sigma^2_v}} d^3 v&=
	N~v^2~dv ~d\phi ~d\cos \theta ~e^{\frac{-3(v^2_E+v^2+2 v_E v \cos\theta)}{2 \sigma^2_v}}\,.\nonumber
	\end{align}
Now carrying out the integration for the angular co-ordinates $\phi$ and $\theta$, we obtain
\begin{align}
	f(v)dv&=N~2\pi v^2
	~dv ~e^{\frac{-3(v^2_E+v^2)}{2 \sigma^2_v}} \int~d\cos \theta ~e^{\frac{-3 v_E v \cos\theta}{ \sigma^2_v}}\nonumber\\& =N~ 2\pi v^2
	~dv ~e^{\frac{-3(v^2_E+v^2)}{2 \sigma^2_v}} \frac{\sigma^2_v}{3 v_E v}\bigg[e^{3\frac{v_E v}{\sigma^2_v}}-e^{-3\frac{v_E v}{\sigma^2_v}}\bigg]\nonumber\\& \simeq N~2 \pi \frac{\sigma^2_v}{3 v_E} v~ dv ~e^{-3 \frac{(v-v_E)^2}{\sigma^2_v}}\nonumber\\ & \equiv A v~ dv ~e^{-3 \frac{(v-v_E)^2}{\sigma^2_v}}
\end{align}
where we have neglected  $e^{-3 \frac{(v+v_E)^2}{\sigma^2_v}}$ compared to $e^{-3 \frac{(v-v_E)^2}{\sigma^2_v}}$ and set $A=N~2 \pi \frac{\sigma^2_v}{3 v_E}$ . 
In Eq.\ref{max_distribution} we identify 
$|\overrightarrow{v_E}|=v_m$, where $v_m$ is the most probable velocity of dark matter.

\bibliographystyle{JHEP}
\bibliography{ref.bib}

		\end{document}